\documentclass[12pt,twoside]{article}   
\usepackage[super,sort,comma]{natbib}

\usepackage{fancyhdr}		

\usepackage{amsmath,amsfonts}
\usepackage{booktabs}
\usepackage{xcolor}

\usepackage{graphicx}
\usepackage{algorithm}
\usepackage{algpseudocode}

\newcommand{\argmin}{\operatornamewithlimits{argmin}}

\usepackage[section]{placeins}   %

\usepackage{graphicx}

\makeatletter \renewcommand\@biblabel[1]{$^{#1}$} \makeatother
\usepackage{hyperref}
\hypersetup{ colorlinks,
    citecolor=blue,
    filecolor=blue,
    linkcolor=blue,
    urlcolor=blue
}

\usepackage{xcolor}

\definecolor{gray}{rgb}{0.6,0.6,0.6}
\definecolor{red}{rgb}{0.85,0,0}
\definecolor{green}{rgb}{0,0.85,0}
\definecolor{blue}{rgb}{0,0,0.85}
\definecolor{beige}{rgb}{0.92,0.87,0.78}
\usepackage[all]{hypcap}    

\begin{document}

\begin{center}
{\sffamily\Large\bfseries Deep Image Prior for Ill-Posed CT Reconstruction With Case-Specific and Patch-Group Priors\par}
\vspace{0.5em}
{\normalsize Ziyu Shu, Xin Qian, Zhixin Pan, Shuang Zhou,\\
Liuxing Shen, Jingxuan Zhu, Zhaozheng Yin, Tiezhi Zhang\par}
\end{center}

\pagenumbering{roman}
\setcounter{page}{1}
\pagestyle{plain}

\begin{abstract}
\noindent\textbf{Background:} Deep image prior (DIP) uses the architecture of an untrained convolutional neural network as an implicit image prior. This enables computed tomography (CT) reconstruction without offline training, but sensitivity to network input and insufficient measurement constraints can cause neural network-specific artifacts in sparse-view and limited-angle settings. Existing methods, such as conditional DIP (CDIP), introduce additional prior information by using reference images as network inputs; however, DIP networks often struggle to effectively remove artifacts already present in those images.

\noindent\textbf{Purpose:} To develop and evaluate prior-information-enhanced DIP (PIE-DIP) for sparse-view and limited-angle CT reconstruction without offline training. The method aims to more effectively incorporate measurement-derived structural information into the network input to guide reconstruction. It also seeks to exploit self-similarity within the reconstruction and selectively use local reference information while limiting the influence of reference-image artifacts.

\noindent\textbf{Methods:} PIE-DIP constructs a case-specific network input from the measured projections rather than using the reference image directly, allowing the linear reconstruction prior of back-projection to complement the nonlinear implicit prior of DIP. We further introduce a collaborative patch-group constraint that provides local guidance through reconstruction self-similarity and, when a reference image is available, position-independent patch matching. During reconstruction, the dictionary progressively shifts from reference-derived to reconstruction-derived patches to reduce the influence of inaccurate reference information.

\noindent Evaluation used the FORBILD phantom and projection data simulated from patient CT images in the Lung Image Database Consortium and Image Database Resource Initiative (LIDC-IDRI). Additional experiments used measured CT data of a carved cheese sample and a walnut from the Finnish Inverse Problems Society. Single-phase few-view reconstruction was further evaluated using Monte Carlo (MC) simulated projections and clinical Varian raw projections from the SPARE Challenge. Comparators included iterative, reference-guided, and DIP-based methods, patch denoising, and pretrained diffusion models. Quality was assessed by signal-to-noise ratio (SNR) and structural similarity index (SSIM).

\noindent\textbf{Results:} Across all evaluated datasets, PIE-DIP achieved the best overall reconstruction performance among the methods compared on each dataset. Experiments also showed that, with appropriate scheduling, this overlap between network optimization and guidance generation introduced no appreciable increase in reconstruction time.

\noindent\textbf{Conclusions:} PIE-DIP improved sparse-view and limited-angle CT reconstruction across simulated and measured datasets without offline training. These results demonstrate the value of combining measurement-derived structural information with selective local patch guidance to exploit useful priors while limiting the influence of inaccurate reference information.
\end{abstract}

\clearpage

\setlength{\baselineskip}{0.7cm}      

\pagenumbering{arabic}
\setcounter{page}{1}
\pagestyle{fancy}
\section{Introduction}
\label{sec:introduction}

Sparse-view and limited-angle computed tomography (CT) are useful when acquisition time, radiation exposure, or scanner geometry is constrained. Limited-angle acquisition can support compact CT systems for interventions, brachytherapy, and non-coplanar external-beam treatment. Sparse-view acquisition can simplify stationary CT systems with fixed X-ray sources~\cite{li2024dual}. It is also relevant to four-dimensional cone-beam CT (4D-CBCT), where respiratory binning leaves few projections available for each phase~\cite{zhang2024review}. The SPARE Challenge provides a public benchmark for sparse-view 4D-CBCT from one-minute acquisitions~\cite{shieh2019spare}. These applications share a reconstruction difficulty: insufficient projections or incomplete angular coverage leave parts of the image weakly constrained.

Filtered back-projection (FBP) is computationally efficient, but its image quality degrades under incomplete acquisition. Iterative reconstruction addresses these conditions by combining the acquisition model with prior knowledge in an optimization framework~\cite{fessler14,willemink2019evolution}. Neural networks offer further approaches, including sinogram preprocessing~\cite{lee2018deep}, end-to-end reconstruction~\cite{zhu2018image}, learned modules within reconstruction frameworks~\cite{shen2019low,shah2018solving}, and image-domain artifact removal~\cite{zhang2016image}. These methods learn useful image priors, but their performance depends on the availability, quality, and representativeness of training data. Generalization, stability, and hallucinated structures remain concerns in medical reconstruction~\cite{wang2018image,antun2020instabilities,bhadra2021hallucinations}.

Deep image prior (DIP) represents the unknown image using an initially untrained convolutional neural network optimized directly against the current measurements, without offline training~\cite{ulyanov2018deep,baguer2020computed,shu2025sdip}. Developments include changes to the architecture and optimization~\cite{gandelsman2019double,chen2020dip}, CT reconstruction variants~\cite{shu2022sparse,shu2024rbp}, explicit regularization such as total variation~\cite{liu2019image} and image-restoration priors~\cite{mataev2019deepred}, and anatomically conditioned PET imaging~\cite{cui2019petdenoising,cui2022unsupervised}. Reconstruction nevertheless remains sensitive to the input, initialization, architecture, and number of iterations~\cite{baguer2020computed,chakrabarty2019spectral,heckel2019denoising,wang2023early}. Under incomplete acquisition, a small projection residual alone may not determine an accurate image.

Although a large training dataset may be unavailable, a small number of related images may be accessible. Examples include earlier scans of the same patient, images of the same object acquired using another modality, or scans of the same anatomical region from other patients. Such reference images can provide prior information without population-scale network training. Reference information can also guide DIP through a pixel-wise image penalty, whose effect depends on reference quality and spatial alignment. Baguer et al.~\cite{baguer2020computed} used image fitting for initialization, then continued with measurement consistency after removing the image-fitting term. This separates initialization from a persistent reference constraint, but does not ensure correction of errors weakly constrained by the measurements.

Conditional DIP (CDIP), introduced for PET denoising with anatomical CT/MR input~\cite{cui2019petdenoising}, supplies this information through the network input. The same input-conditioning strategy can be adapted to CT reconstruction. However, direct input substitution does not distinguish useful reference structures from mismatched content or existing artifacts. These artifacts may therefore remain in the reconstruction, particularly when the available measurements provide little information for correcting them.

Local image structures offer a more selective source of information. Block-matching and three-dimensional filtering (BM3D) groups similar patches and performs collaborative transform-domain filtering~\cite{dabov2007image}. Shared structures can be retained while components inconsistent across matched patches are attenuated. This is useful for incomplete CT, where missing views can produce streaks across large image regions. Extending the patch pool to include an imperfect reference can provide additional local structures without requiring pixel-wise correspondence. A mixed dictionary can then combine these structures with self-similarity in the current reconstruction.

These observations motivate two complementary improvements to DIP: introducing structural priors through the network input while reducing the influence of artifacts, and selectively extracting useful information from reference images that may contain mismatched structures or artifacts. We propose prior-information-enhanced DIP (PIE-DIP) to combine these improvements without offline training. The input component introduces a case-specific linear reconstruction prior through modified back-projection, complementing the nonlinear implicit prior of DIP. The patch component uses self-similarity within the current reconstruction and, when a reference image is available, extracts useful local structures without requiring pixel-wise correspondence between the reference and target images. Together, these components improve reconstruction from incomplete measurements.

The main contributions are as follows:
\begin{itemize}
\item We construct the network input directly from the measured projections, allowing the linear prior and object-specific structural information provided by back-projection to complement the nonlinear implicit prior of DIP. This input requires no external reference image; an available reference can further modulate the input.

\item We introduce a collaborative patch-group prior that guides reconstruction using a mixed dictionary of reconstruction and optional reference patches. During optimization, the dictionary progressively shifts from reference-derived to reconstruction-derived patches, reducing reliance on the reference. With appropriate scheduling, the next dictionary and guiding patches are generated in parallel with network optimization, introducing no appreciable increase in reconstruction time.

\item We assess the two components through ablation studies and evaluate PIE-DIP using the FORBILD phantom~\cite{yu2012simulation}, LIDC-IDRI dataset~\cite{armato2011lung}, measured cheese and walnut CT data from the Finnish Inverse Problems Society~\cite{hamalainen2015tomographic,bubba2017tomographic}, as well as single-phase Monte Carlo (MC) simulated projections and clinical Varian raw projections from SPARE~\cite{shieh2019spare}. Comparisons include iterative reconstruction, conventional reference-constrained methods, patch-based methods, DIP variants, and task-specific pretrained diffusion models. PIE-DIP achieves the best overall reconstruction performance among the methods compared on each dataset.
\end{itemize}

Section~\ref{sec:Proposed Methods} presents the reconstruction framework and PIE-DIP components. Section~\ref{sec:exp} describes the experiments and results. Sections~\ref{sec:diss} and~\ref{sec:conclusion} discuss the findings and conclude the paper.

\section{Proposed Methods}\label{sec:Proposed Methods}
PIE-DIP combines a measurement-derived network input with a collaborative patch-group prior. The modified back-projection input introduces structural information from the current measurements. The patch-group prior uses self-similarity in the evolving reconstruction and incorporates local reference information when available.

\subsection{Reconstruction Framework}\label{sec:framework}
DIP represents the unknown image by the output of an initially untrained convolutional neural network (CNN). For a linear imaging system, the network parameters are optimized against the measured projections:
\begin{equation}
    \boldsymbol w^* = \argmin_{\boldsymbol w}  ||\boldsymbol y - \boldsymbol{\rm H} G(\boldsymbol w|\boldsymbol z)||^2_2, \quad
\boldsymbol x^* = G(\boldsymbol w^*|\boldsymbol z),
\label{eq:dip}
\end{equation}
where \(\boldsymbol y\) denotes the measured projections, \(\boldsymbol{\rm H}\) is the imaging operator, and \(G(\boldsymbol w|\boldsymbol z)\) is the CNN with parameters \(\boldsymbol w\) and fixed input \(\boldsymbol z\). The final image \(\boldsymbol x^*\) is the optimized network output. Conventional DIP uses fixed random noise as its input~\cite{baguer2020computed,shu2022sparse}. Using CDIP input conditioning for CT reconstruction gives the same objective with \(\boldsymbol z=\boldsymbol x_{\rm ref}\), where \(\boldsymbol x_{\rm ref}\) is a reference image~\cite{cui2019petdenoising}. PIE-DIP constructs a measurement-derived input and adds a collaborative patch-group term, as described below.

\subsection{Modified Back-Projection Input}\label{sec:modifiedinput}
We first consider the case in which no additional reference image is available. A natural option is to use an image produced by another reconstruction method, such as iterative reconstruction (IR) or DIP, as the input to CDIP, since such images can have relatively high reconstruction accuracy. However, these images are obtained by optimizing objectives that include measurement consistency. They have already been optimized to reduce the projection residual and may lie near low-loss solutions of the data-fitting problem. A sufficiently flexible DIP network can then quickly approach an approximate identity mapping, leaving limited scope for its implicit prior to suppress the remaining artifacts.

In contrast, a single back-projection image \(\boldsymbol x_{\rm BP}\) provides coarse object-dependent structural information and a linear reconstruction prior without iterative data fitting. This linear prior complements the nonlinear implicit prior of DIP. Moreover, its data-fitting objective value remains well above the minimum, leaving substantial scope for further DIP optimization. These properties motivate using back-projection as the network input prior.

Raw back-projection can nevertheless contain coherent streaks caused by missing views. We therefore further modulate it with random noise, with the aim of weakening the influence of these streak patterns during network optimization and suppressing their propagation to the reconstruction. The back-projection remains the source of object-specific structure, while random modulation is intended to reduce reliance on its unmodified artifact pattern.

\begin{figure*}
	\centering
	\begin{minipage}[b]{.32\linewidth}
		\centering
		\centerline{\includegraphics[width=\linewidth]{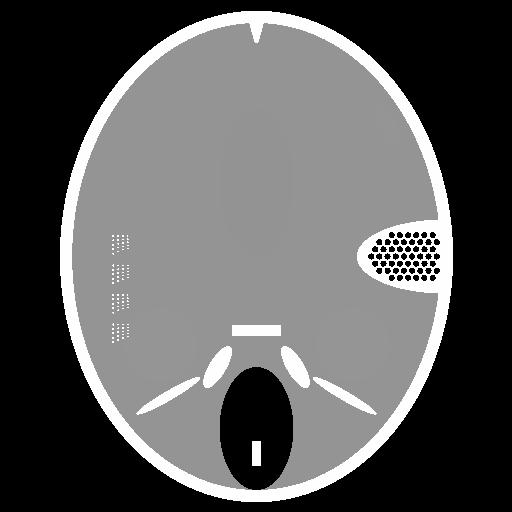}}
		  \vspace{0.5cm}
		\centerline{Ground Truth}\medskip
	\end{minipage}
	\begin{minipage}[b]{.16\linewidth}
		\begin{minipage}[b]{1\linewidth}
		\centering
		\centerline{\includegraphics[width=\linewidth]{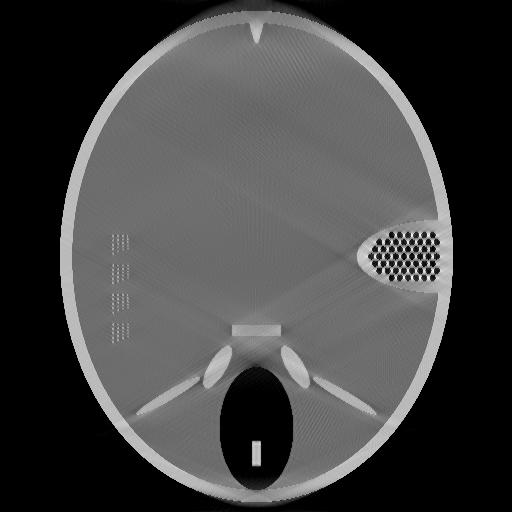}}
		\centerline{\tiny IR 19.59~dB}\medskip
	\end{minipage}
 
    \begin{minipage}[b]{1\linewidth}
		\centering
		\centerline{\includegraphics[width=\linewidth]{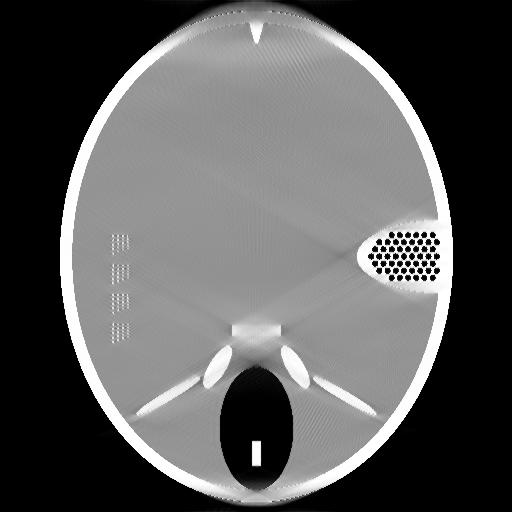}}
		\centerline{\tiny CDIP-IR 19.25~dB}\medskip
	\end{minipage}
	\end{minipage}
\begin{minipage}[b]{.16\linewidth}
		\begin{minipage}[b]{1\linewidth}
		\centering
		\centerline{\includegraphics[width=\linewidth]{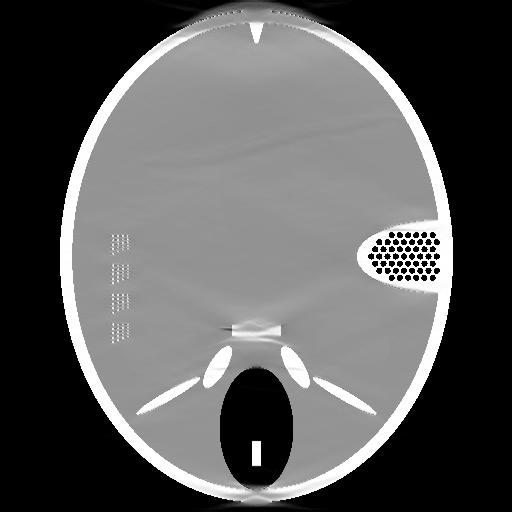}}
		\centerline{\tiny DIP 20.36~dB}\medskip
	\end{minipage}
 
    \begin{minipage}[b]{1\linewidth}
		\centering
		\centerline{\includegraphics[width=\linewidth]{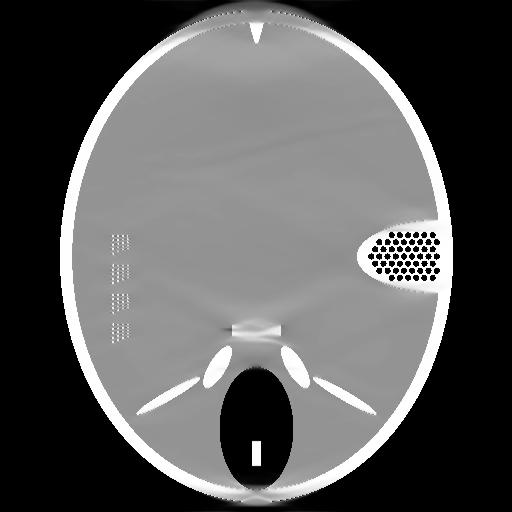}}
		\centerline{\tiny CDIP-DIP 20.07~dB}\medskip
	\end{minipage}
	\end{minipage}
\begin{minipage}[b]{.16\linewidth}
		\begin{minipage}[b]{1\linewidth}
		\centering
		\centerline{\includegraphics[width=\linewidth]{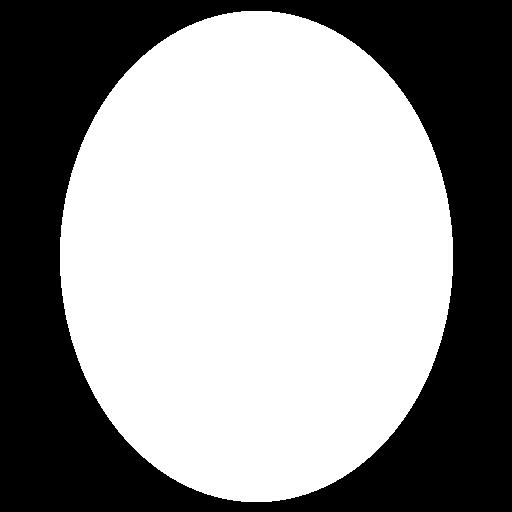}}
		\centerline{\tiny BINARY MASK}\medskip
	\end{minipage}
 
    \begin{minipage}[b]{1\linewidth}
		\centering
		\centerline{\includegraphics[width=\linewidth]{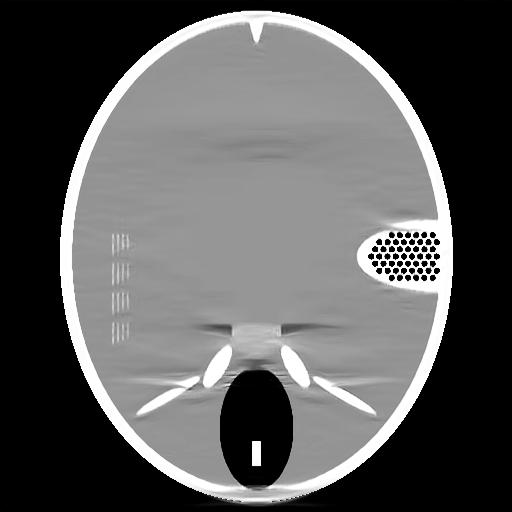}}
		\centerline{\tiny CDIP-MASK 22.34~dB}\medskip
	\end{minipage}
	\end{minipage}   
	\begin{minipage}[b]{.16\linewidth}
		\begin{minipage}[b]{1\linewidth}
		\centering
		\centerline{\includegraphics[width=\linewidth]{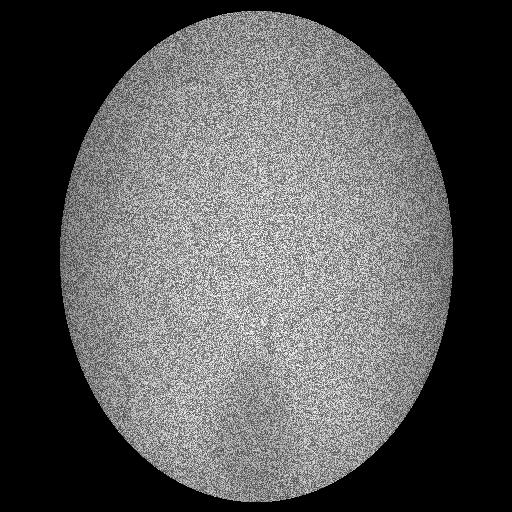}}
		\centerline{\tiny PIE-DIP INPUT}\medskip
	\end{minipage}
 
    \begin{minipage}[b]{1\linewidth}
		\centering
		\centerline{\includegraphics[width=\linewidth]{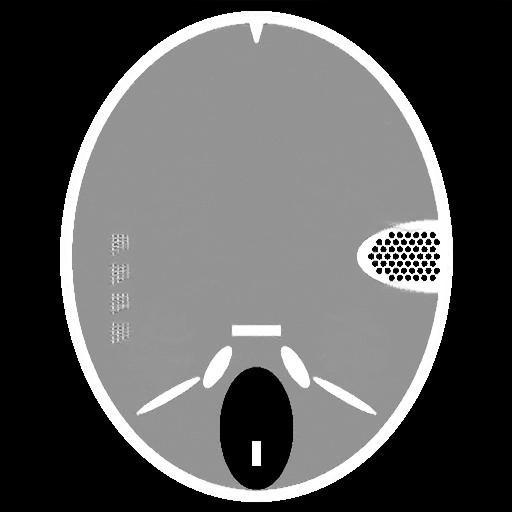}}
		\centerline{\tiny PIE-DIP 30.25~dB}\medskip
	\end{minipage}
	\end{minipage}

	\caption{
    Influence of the fixed network input on limited-angle CT reconstruction of the FORBILD phantom~\cite{yu2012simulation} using 120 evenly distributed views from $0^\circ$ to $120^\circ$. The leftmost image is the ground truth. In each remaining column, the upper image is the network input and the lower image is the corresponding DIP reconstruction. The tested inputs are an IR reconstruction, a DIP reconstruction, a binary reference image, and the proposed modified back-projection input. The final column evaluates the modified back-projection input alone, without the patch-group prior. The reported values are SNR in dB.}
	\label{fig:cdip}
\end{figure*}

Fig.~\ref{fig:cdip} illustrates the motivation. When an iterative reconstruction (IR) or DIP reconstruction is used as the CDIP input, the output remains close to that input and retains its artifacts. A binary reference provides coarse boundary information, but network-specific artifacts remain visible. In contrast, the modified back-projection input improves the reconstruction in this example.

Let \(\boldsymbol y\) denote the measured projections. We multiply the back-projection \(\boldsymbol x_{\rm BP}=\boldsymbol {\rm H}^{\rm T}\boldsymbol y\) elementwise by fixed uniform random numbers \(\boldsymbol\epsilon\) in \([0,1]\), then normalize the result by its Euclidean norm. The resulting input is
\begin{equation}
    \boldsymbol z
    =
    {\cal N}_2\left(
    \boldsymbol{\rm H}^{\rm T}\boldsymbol y
    \odot\boldsymbol\epsilon
    \right),
\label{eq:modified_bp_input}
\end{equation}
where \(\boldsymbol{\rm H}^{\rm T}\) is the back-projection operator, \(\odot\) denotes elementwise multiplication, and \({\cal N}_2(\boldsymbol v)=\boldsymbol v/\|\boldsymbol v\|_2\). Normalization controls the overall input scale. The random field is sampled once, and the resulting input is held fixed during network optimization.

\begin{figure*}
	\centering
	\includegraphics[width=0.75\linewidth]{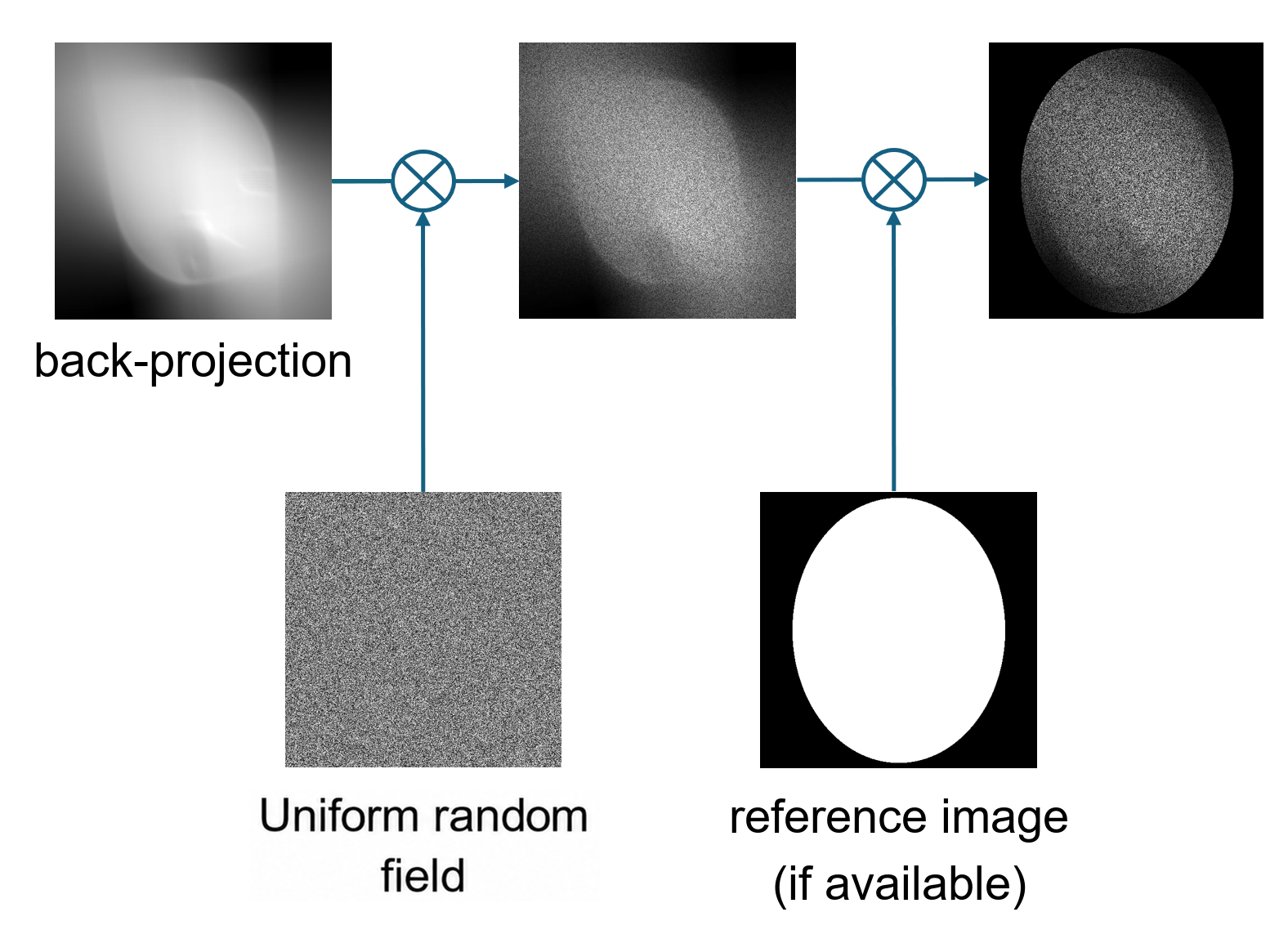}
    \caption{PIE-DIP input construction for a limited-angle CT example. The back-projection is multiplied by a fixed uniform random field. An optional reference image, illustrated here by a binary reference image, can be included as an additional multiplicative modulation. The final product is normalized by its Euclidean norm before being supplied to the DIP network.}
	\label{ini_input}
\end{figure*}

This measurement-derived input construction is compatible with the reference-conditioning principle of conventional CDIP. When a high-quality reference image \(\boldsymbol x_{\rm ref}\) is available, it can provide an additional multiplicative modulation:
\begin{equation}
\boldsymbol z
=
{\cal N}_2\left(
\boldsymbol{\rm H}^{\rm T}\boldsymbol y
\odot\boldsymbol\epsilon
\odot\boldsymbol x_{\rm ref}
\right).
\label{eq:modified_bp_input_ref}
\end{equation}
This optional modulation does not impose the reference as a pixel-wise reconstruction target. The reference can also supply patches for the collaborative prior in Section~\ref{sec:patchprior}. Fig.~\ref{ini_input} summarizes input construction.

\subsection{Collaborative Patch-Group Prior}\label{sec:patchprior}
The patch-group prior uses repeated local structures in the reconstruction and, when available, reference images. Patches are matched by similarity rather than spatial location, so pixel-wise correspondence between the reference and target images is not required. Matched patches are grouped and collaboratively filtered to form guiding targets. Collaborative filtering can preserve common local structures while attenuating artifact components that are inconsistent within the group, helping suppress directional streak artifacts caused by missing views.

Let \(\boldsymbol x_{\rm ref}\) denote the reference image and \(\boldsymbol x_i=G(\boldsymbol w_i|\boldsymbol z)\) the reconstruction snapshot saved when guidance generation is launched at iteration \(i\). Overlapping \(s\times s\) patches are extracted with stride \(r\). Each vectorized patch \(\boldsymbol p\in\mathbb R^{s^2}\) is mean-centered and normalized using a stabilized Euclidean norm:
\begin{equation}
    \phi(\boldsymbol p)=
    \frac{\boldsymbol p-\mu(\boldsymbol p)\boldsymbol 1}
    {\|\boldsymbol p-\mu(\boldsymbol p)\boldsymbol 1\|_2+\varepsilon},
    \qquad
    \mu(\boldsymbol p)=\frac{1}{s^2}\boldsymbol 1^{\rm T}\boldsymbol p,
\label{eq:patch_normalization}
\end{equation}
Here, \(\varepsilon\) is a small constant for numerical stability. The normalized patch sets are denoted by \({\cal S}_{\rm ref}\) and \({\cal S}_i\). A dictionary of \(D\) patches is sampled without replacement from these sets:
\begin{equation}
{\cal D}_i =
\operatorname{Sample}({\cal S}_{\rm ref},DP_i)
\cup
\operatorname{Sample}({\cal S}_i,D(1-P_i)).
\label{eq:mixed_dictionary}
\end{equation}
The reference fraction is \(P_i\), and the reconstruction fraction is \(1-P_i\). Sample counts are rounded to integers while keeping the total equal to \(D\).

When a reference image is available, \(P_i\) follows a predefined sigmoid schedule:
\begin{equation}
    P_i = \frac{1}{1+\exp[(i-i_c)/\tau]},
    \label{eq:sampling_schedule}
\end{equation}
The parameters \(i_c\) and \(\tau\) control the transition center and width. The fraction \(P_i\) is evaluated at the task's launch iteration \(i\). Dictionary construction and query selection both use the snapshot \(\boldsymbol x_i\). Early dictionaries therefore draw mainly from the reference image, while later dictionaries rely increasingly on the reconstruction. This transition reduces dependence on the reference during later refinement. Without an external reference, \(P_i=0\).

For a query patch \(\boldsymbol q\), let \(\widehat{\boldsymbol q}=\phi(\boldsymbol q)\). Similarity is measured by the squared Euclidean distance between the normalized query and each dictionary patch \(\boldsymbol d_j\in{\cal D}_i\):
\begin{equation}
    d(\widehat{\boldsymbol q},\boldsymbol d_j)
    =\|\widehat{\boldsymbol q}-\boldsymbol d_j\|_2^2.
\label{eq:patch_distance}
\end{equation}
Let \(j_1,\ldots,j_N\) index the \(N\) dictionary patches with the smallest distances. The query and these nearest neighbors are stacked row-wise:
\begin{equation}
    \boldsymbol G_{\boldsymbol q}
    =
    [\widehat{\boldsymbol q},\boldsymbol d_{j_1},\ldots,
    \boldsymbol d_{j_N}]^{\rm T}
    \in\mathbb R^{(N+1)\times s^2}.
\label{eq:patch_group}
\end{equation}
Each group contains one query and \(N\) matched dictionary patches. The dictionary is searched in chunks to reduce peak GPU memory while retaining the globally nearest \(N\) matches.

Collaborative filtering applies a one-dimensional discrete cosine transform (DCT) along the group dimension, followed by hard thresholding and an inverse DCT:
\begin{equation}
    \widetilde{\boldsymbol G}_{\boldsymbol q}
    =\boldsymbol C_{N+1}^{\rm T}
    {\cal T}_{\eta_{\rm cf}}
    (\boldsymbol C_{N+1}\boldsymbol G_{\boldsymbol q}),
\label{eq:group_filter}
\end{equation}
Here, \(\boldsymbol C_{N+1}\) is an orthonormal DCT matrix. The operator \({\cal T}_{\eta_{\rm cf}}\) sets coefficients with magnitudes no greater than \(\eta_{\rm cf}\) to zero and leaves the others unchanged. The first row of the filtered group corresponds to the query and provides its guiding patch:
\begin{equation}
    \widetilde{\boldsymbol q}
    =\left(\widetilde{\boldsymbol G}_{\boldsymbol q}\right)_{1,:}.
\label{eq:guiding_patch}
\end{equation}
This filtering follows the nonlocal grouping principle of BM3D~\cite{dabov2007image}, but differs from standard BM3D in two ways. Filtering acts only along the group dimension, without a two-dimensional transform within individual patches. It aims to suppress differences across similar patches without an explicit within-patch spatial transform. When a reference is available, the dictionary also combines reference and reconstruction patches according to the schedule above, rather than drawing patches only from the current image.

For the active dictionary generated at iteration \(i\), let \(I_i\) denote the selected query locations and \(\widetilde{\boldsymbol q}^{[i]}_\ell\) the corresponding guiding patches. With \(E_\ell\) denoting patch extraction at location \(\ell\), the patch-group loss compares the current network output with these targets:
\begin{equation}
    {\cal L}_{\rm cp}(\boldsymbol w)
    =\frac{1}{|I_i|s^2}
    \sum_{\ell\in I_i}
    \left\|
    \phi\!\left(E_{\ell}G(\boldsymbol w|\boldsymbol z)\right)
    -\widetilde{\boldsymbol q}^{[i]}_{\ell}
    \right\|_2^2.
\label{eq:patch_group_loss}
\end{equation}
Gradients pass through the current network output and its patch normalization. The guiding targets and their locations are treated as constants during back-propagation; matching and filtering are performed only when the dictionary is updated.

With an external reference, early dictionaries mainly contain fixed reference patches, so the guiding patches change slowly. Later, dictionaries mainly contain reconstruction patches, but the network output has become relatively stable, again limiting changes in the guidance. This motivates periodic updates rather than regenerating the guiding patches at every DIP iteration.

At iteration \(i\), patches \({\cal S}_i\) are extracted from \(\boldsymbol x_i\) and combined with reference patches when available to construct \({\cal D}_i\) and finally generate the guiding patches \(\widetilde{\boldsymbol q}^{[i]}_\ell\), while network optimization continues in parallel. The new dictionary and guiding patches become active at \(i+T\), where \(T\) is the update interval. If generation takes less time than \(T\) DIP iterations, this update introduces no additional waiting time in reconstruction.

\begin{figure}[htpb]
    \centering
     \includegraphics[width=\linewidth]{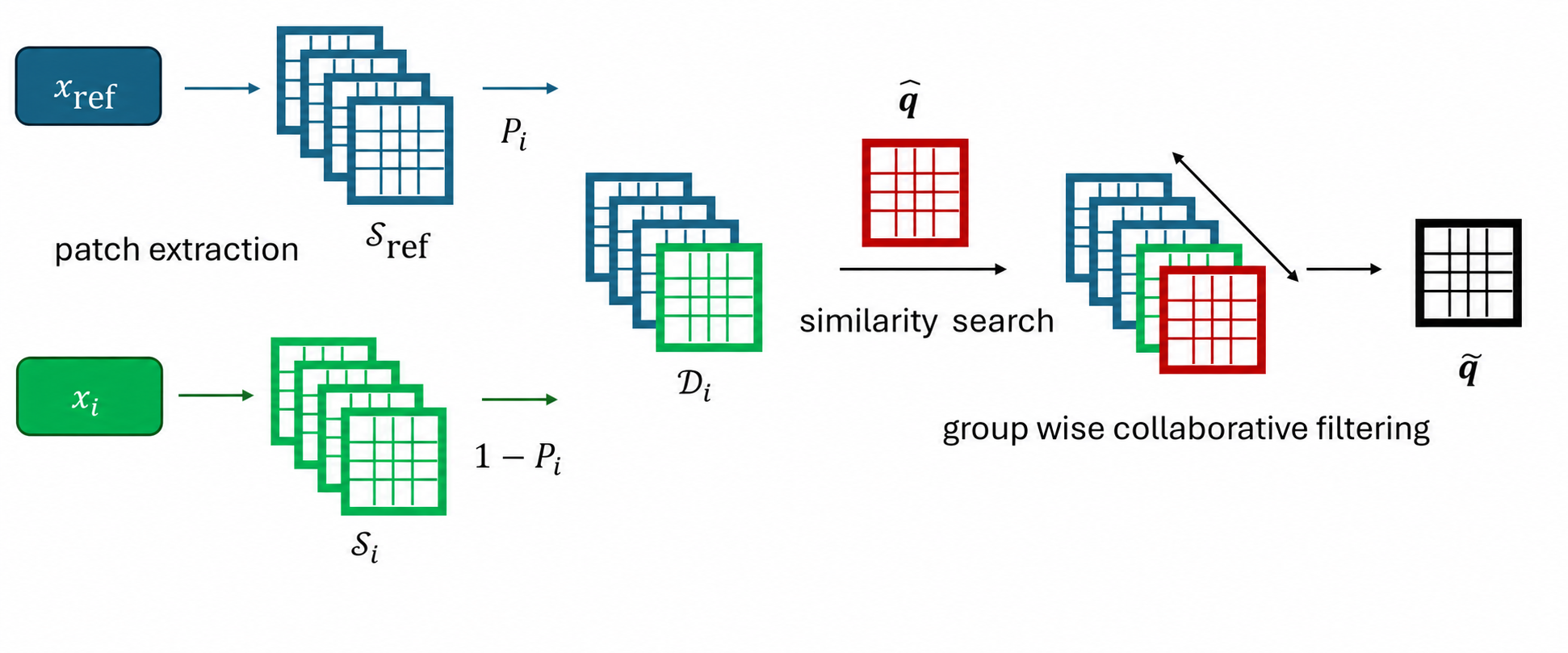}
    \caption{Workflow of the proposed group-wise collaborative filtering prior. Normalized patch sets \({\cal S}_{\rm ref}\) and \({\cal S}_i\) are extracted from the reference image \(\boldsymbol x_{\rm ref}\) and a saved reconstruction snapshot \(\boldsymbol x_i\). A mixed dictionary \({\cal D}_i\) is constructed using proportions \(P_i\) and \(1-P_i\), with 15,876 patches under the default settings. Each normalized query \(\widehat{\boldsymbol q}\) is grouped with its 16 nearest dictionary patches, and group-wise filtering produces the guiding patch \(\widetilde{\boldsymbol q}\). The network uses guiding patches from the current dictionary for optimization while the next dictionary and guiding patches are generated in parallel. The dictionary \({\cal D}_i\) and its guiding patches become active at \(i+T\). If generation finishes within the time required for \(T\) DIP iterations, the update adds no waiting time.}
    \label{fig:groupwise_cf}
\end{figure}

Fig.~\ref{fig:groupwise_cf} shows the workflow.

\subsection{Overall PIE-DIP Reconstruction}
PIE-DIP combines the modified input in Section~\ref{sec:modifiedinput} with the patch-group prior in Section~\ref{sec:patchprior}. For a fixed input \(\boldsymbol z\) and the currently active guiding patches, the network-update objective is
\begin{equation}
    \boldsymbol w^*
    =
    \argmin_{\boldsymbol w}
    \left\|
    \boldsymbol y - \boldsymbol{\rm H}G(\boldsymbol w|\boldsymbol z)
    \right\|_2^2
    +
    \lambda_{\rm cp}
    {\cal L}_{\rm cp}(\boldsymbol w),
    \qquad
    \boldsymbol x^*=G(\boldsymbol w^*|\boldsymbol z),
\label{eq:piedip_objective}
\end{equation}
where \({\cal L}_{\rm cp}(\boldsymbol w)\) is defined in Eq.~\eqref{eq:patch_group_loss}, \(\lambda_{\rm cp}\) controls its contribution, and \(\boldsymbol x^*\) is the reconstructed image. RMSProp updates the network parameters using the guiding patches from the active dictionary.

The two components act at different levels: the modified input supplies measurement-derived structure, while the patch term adds local self-similarity and reference guidance.

Given the strong fitting capacity of DIP networks demonstrated in previous DIP and CDIP studies~\cite{ulyanov2018deep,cui2022unsupervised}, we do not expect the modified input and additional soft regularization to hinder convergence under the optimization settings used here.

Algorithm~\ref{alg:collaborative_patch_filtering} summarizes the overall PIE-DIP reconstruction procedure, including input construction, network optimization, and parallel guidance updates.

\begin{algorithm}[htbp]
    \caption{PIE-DIP}
    \label{alg:collaborative_patch_filtering}
    \hspace*{0.02in}{\bf Input:}
    measurement model \((\boldsymbol H,\boldsymbol y)\), network \(G\), optional reference image \(\boldsymbol x_{\rm ref}\), iteration count \(N_{\rm iter}\), update interval \(T\), and parameters \(s,r,D,N,i_c,\tau,\eta_{\rm cf},\lambda_{\rm cp}\).
    \begin{algorithmic}[1]
        \State Compute the back-projection \(\boldsymbol x_{\rm BP}=\boldsymbol H^{\rm T}\boldsymbol y\).
        \State Sample fixed uniform random numbers \(\boldsymbol\epsilon\) in \([0,1]\) and set \(\boldsymbol z=\boldsymbol x_{\rm BP}\odot\boldsymbol\epsilon\).
        \If{a reference image is available}
            \State Set \(\boldsymbol z\gets\boldsymbol z\odot\boldsymbol x_{\rm ref}\).
        \EndIf
        \State Normalize \(\boldsymbol z\gets{\cal N}_2(\boldsymbol z)\) and keep it fixed throughout optimization.
        \State Initialize \(\boldsymbol w_1\); no guiding patches are active initially.
        \For{\(i=1,\ldots,N_{\rm iter}\)}
            \State Compute \(\boldsymbol x_i=G(\boldsymbol w_i\mid\boldsymbol z)\).
            \If{\(i\bmod T=0\)}
                \State Launch \Call{GenerateGuidance}{}\((\boldsymbol x_i,i)\) in the background using a fixed snapshot of \(\boldsymbol x_i\); schedule its output for activation at iteration \(i+T\).
            \EndIf
            \State Compute \({\cal L}_{\rm cp}\) using the active guiding patches and Eq.~\eqref{eq:patch_group_loss}; set \({\cal L}_{\rm cp}=0\) if none are active.
            \State Perform one RMSProp step using Eq.~\eqref{eq:piedip_objective} to obtain \(\boldsymbol w_{i+1}\).
        \EndFor
        \State {\bf Output:} \(\boldsymbol x^*=G(\boldsymbol w_{N_{\rm iter}+1}\mid\boldsymbol z)\).
        \Statex
        \Function{GenerateGuidance}{}\((\boldsymbol x_i,i)\)
            \State Extract and normalize patches from \(\boldsymbol x_i\) and, if available, \(\boldsymbol x_{\rm ref}\) using Eq.~\eqref{eq:patch_normalization}, obtaining \({\cal S}_i\) and optionally \({\cal S}_{\rm ref}\).
            \State If a reference is available, evaluate \(P_i\) using Eq.~\eqref{eq:sampling_schedule} and construct \({\cal D}_i\) using Eq.~\eqref{eq:mixed_dictionary}; otherwise, set \(P_i=0\) and sample \({\cal D}_i\) from \({\cal S}_i\).
            \State Select query locations \(I_i\) from the saved snapshot.
            \For{each \(\ell\in I_i\)}
                \State Normalize the query as \(\widehat{\boldsymbol q}=\phi(E_\ell\boldsymbol x_i)\).
                \State Find its \(N\) nearest dictionary patches using Eq.~\eqref{eq:patch_distance} and form the group using Eq.~\eqref{eq:patch_group}.
                \State Apply the group-wise DCT, hard thresholding, and inverse DCT using Eq.~\eqref{eq:group_filter}.
                \State Save the first filtered row as \(\widetilde{\boldsymbol q}^{[i]}_\ell\) using Eq.~\eqref{eq:guiding_patch}.
            \EndFor
            \State \Return \(\{\widetilde{\boldsymbol q}^{[i]}_\ell\}_{\ell\in I_i},I_i\).
        \EndFunction
    \end{algorithmic}
\end{algorithm}

\subsection{Computational Cost}
\label{sec:complex}
Input construction requires one back-projection and several elementwise operations before optimization, so this cost is incurred only once. During optimization, the network uses guiding targets from the current dictionary while the next dictionary and guiding patches are generated in parallel, aiming to avoid additional reconstruction time. Query generation, dictionary construction, matching, and filtering are performed once per update cycle. Ordinary iterations evaluate the patch loss against the active guiding targets without repeating these operations. Dictionary search is divided into chunks to limit peak GPU memory.

Timing was averaged over the complete 5000-iteration reconstruction and includes periodic dictionary construction and guidance updates. On an NVIDIA GeForce RTX 4090 GPU, the average wall-clock time per network iteration was 0.0646~s without collaborative filtering and 0.0650~s with the proposed module, an increase of 0.0004~s, or approximately 0.62\%.

For FP32 computation with one \(512\times512\) image, the default \(15{,}876\times144\) patch dictionary occupies approximately 8.7~MiB, whereas DIP without the patch-group module uses approximately 2.4--3.2~GiB of peak allocated GPU memory. The dictionary therefore represents less than 0.4\% of the memory used by the DIP reconstruction itself; temporary tensors used during chunked patch search and collaborative filtering are allocated only during guidance updates.

\section{Experiments and Results}\label{sec:exp}
Experiments use the FORBILD phantom~\cite{yu2012simulation}, LIDC-IDRI patient-image simulations~\cite{armato2011lung}, measured CT data from the Finnish Inverse Problems Society~\cite{hamalainen2015tomographic,bubba2017tomographic}, and the SPARE Challenge dataset~\cite{shieh2019spare}.

Unless otherwise specified, reconstructions use fan-beam geometry and \(512\times512\) images. Forward and back-projections use ray tracing with detector blurring. FORBILD and LIDC-IDRI projections are simulated from ground-truth images upsampled to \(1024\times1024\) to reduce inverse-crime effects. Implementations using lookup tables~\cite{ha2017look}, box splines~\cite{zhang2019box}, and separable footprints~\cite{long20103d} yield consistent comparative trends.

ASD-POCS serves as the iterative reconstruction (IR) baseline. Unless otherwise specified, ASD-POCS and all DIP-based methods run for 5000 iterations. DIP methods share the same U-Net backbone, including in SPARE experiments. DIP networks are implemented in PyTorch with default initialization and optimized using RMSProp with an initial learning rate of \(1\times10^{-4}\), multiplied by 0.9 every 250 iterations.

The default patch settings are \(\lambda_{\rm cp}=0.25\), patch size \(12\times12\), stride 4, and dictionary size \(D=15{,}876\). Normalized dictionary patches are sampled without replacement from the reference and reconstruction pools in proportions \(P_i\) and \(1-P_i\). Each update selects 512 query patches from a saved reconstruction snapshot and matches each to its 16 nearest dictionary neighbors. Groups are filtered only along the group dimension using a one-dimensional DCT and hard thresholding with \(\eta_{\rm cf}=0.05\). The dictionary and guiding patches are generated in parallel with network optimization; results from tasks launched at \(i\) become active at \(i+T\). The update interval is \(T=100\) iterations.

When a reference is used, the mixing schedule uses \(i_c=2500\) and \(\tau=400\). Hyperparameters were selected once on an independent LIDC-IDRI validation subset and fixed for all test cases.

\subsection{Ablation Experiments}\label{sec:ablation}\label{sec:patchpriorexp}
We evaluate the modified back-projection input and patch-group prior separately, then examine their effects on optimization. All tests use LIDC-IDRI images and show consistent trends under sparse-view and limited-angle conditions. Representative limited-angle results are presented for the input ablation and sparse-view results for the patch-prior ablation.

\subsubsection{Effect of the Modified Back-Projection Input}
\begin{figure*}
    \centering
    \begin{minipage}[b]{.19\linewidth}
        \centering
        \centerline{\includegraphics[width=\linewidth]{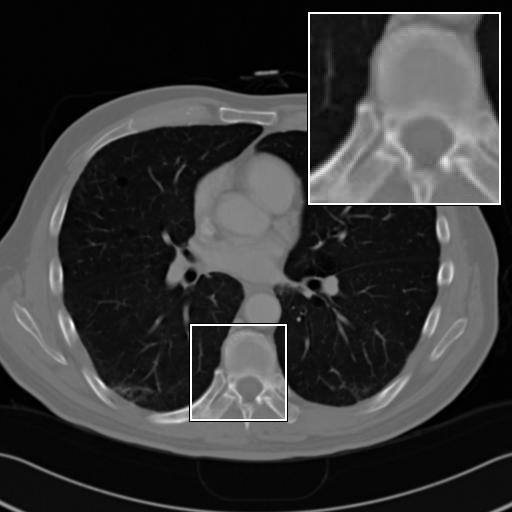}}
        \centerline{\scriptsize\shortstack{(a) Ground truth\\-/-}}\medskip
    \end{minipage}
    \begin{minipage}[b]{.19\linewidth}
        \centering
        \centerline{\includegraphics[width=\linewidth]{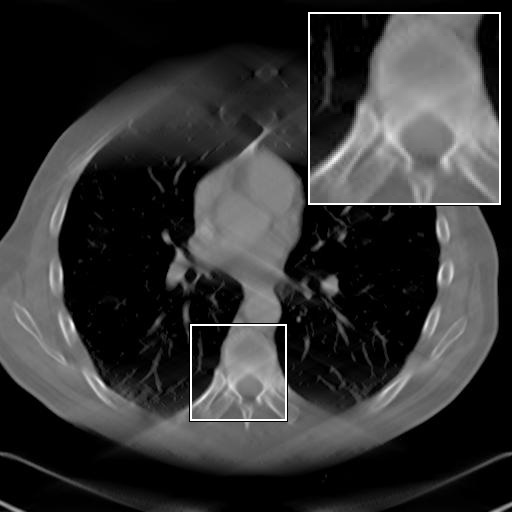}}
        \centerline{\scriptsize\shortstack{(b) ASD-POCS\\13.80/0.861}}\medskip
    \end{minipage}
    \begin{minipage}[b]{.19\linewidth}
        \centering
        \centerline{\includegraphics[width=\linewidth]{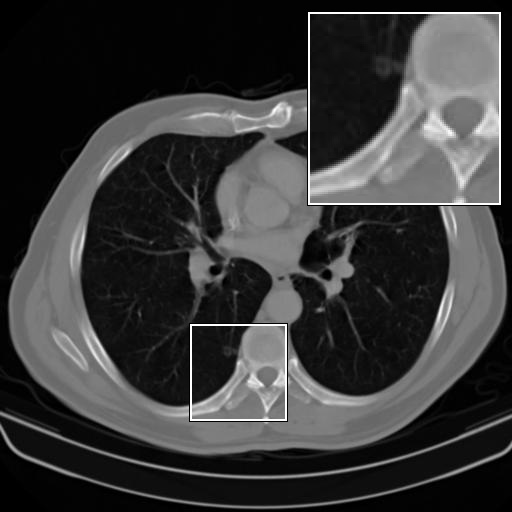}}
        \centerline{\scriptsize\shortstack{(c) Reference 1\\-/-}}\medskip
    \end{minipage}
    \begin{minipage}[b]{.19\linewidth}
        \centering
        \centerline{\includegraphics[width=\linewidth]{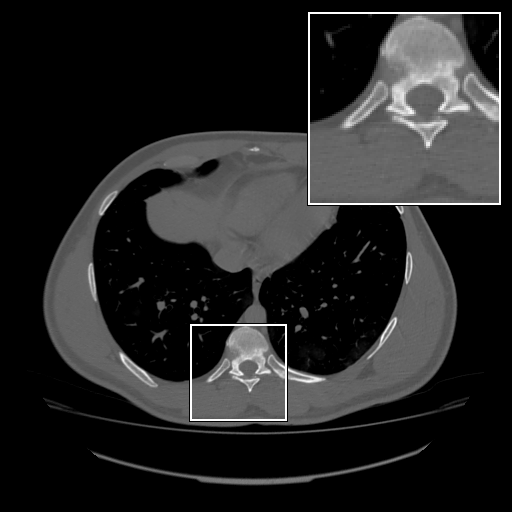}}
        \centerline{\scriptsize\shortstack{(d) Reference 2\\-/-}}\medskip
    \end{minipage}

    \begin{minipage}[b]{.19\linewidth}
        \centering
        \centerline{\includegraphics[width=\linewidth]{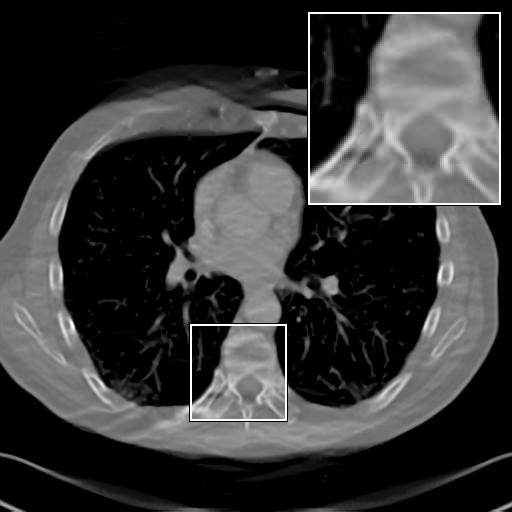}}
        \centerline{\scriptsize\shortstack{(e) DIP\\13.99/0.856}}\medskip
    \end{minipage}
    \begin{minipage}[b]{.19\linewidth}
        \centering
        \centerline{\includegraphics[width=\linewidth]{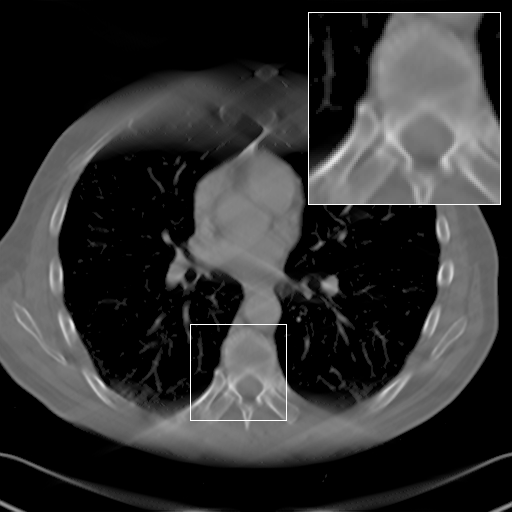}}
        \centerline{\scriptsize\shortstack{(f) CDIP + ASD\\13.45/0.864}}\medskip
    \end{minipage}
    \begin{minipage}[b]{.19\linewidth}
        \centering
        \centerline{\includegraphics[width=\linewidth]{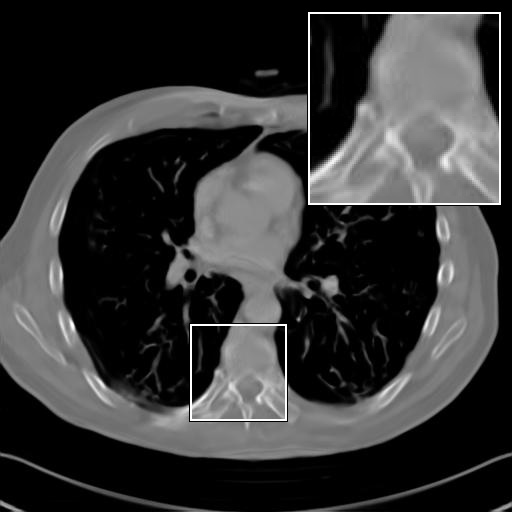}}
        \centerline{\scriptsize\shortstack{(g) CDIP + Ref. 1\\19.19/0.897}}\medskip
    \end{minipage}
    \begin{minipage}[b]{.19\linewidth}
        \centering
        \centerline{\includegraphics[width=\linewidth]{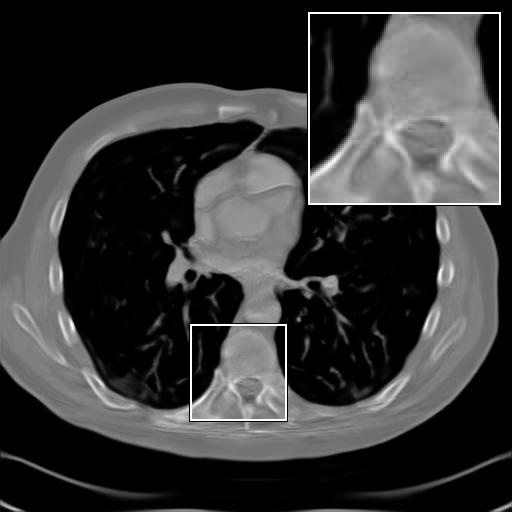}}
        \centerline{\scriptsize\shortstack{(h) CDIP + Ref. 2\\16.02/0.853}}\medskip
    \end{minipage}

    \begin{minipage}[b]{.19\linewidth}
        \centering
        \centerline{\includegraphics[width=\linewidth]{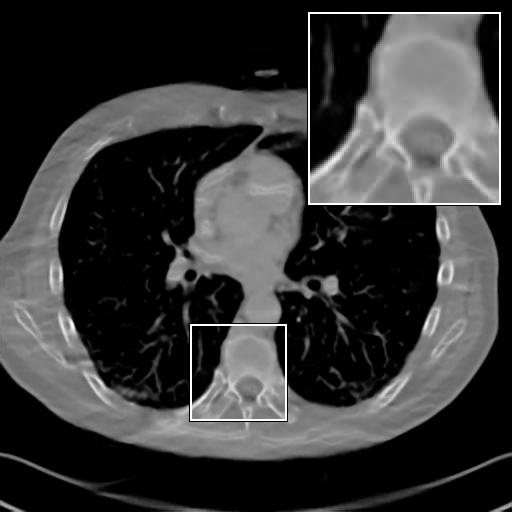}}
        \centerline{\scriptsize\shortstack{(i) Mod. BP\\17.27/0.879}}\medskip
    \end{minipage}
    \begin{minipage}[b]{.19\linewidth}
        \centering
        \centerline{\includegraphics[width=\linewidth]{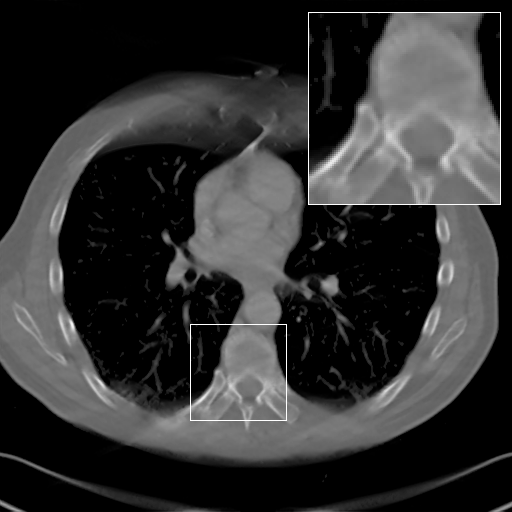}}
        \centerline{\scriptsize\shortstack{(j) Mod. BP + ASD\\14.14/0.875}}\medskip
    \end{minipage}
    \begin{minipage}[b]{.19\linewidth}
        \centering
        \centerline{\includegraphics[width=\linewidth]{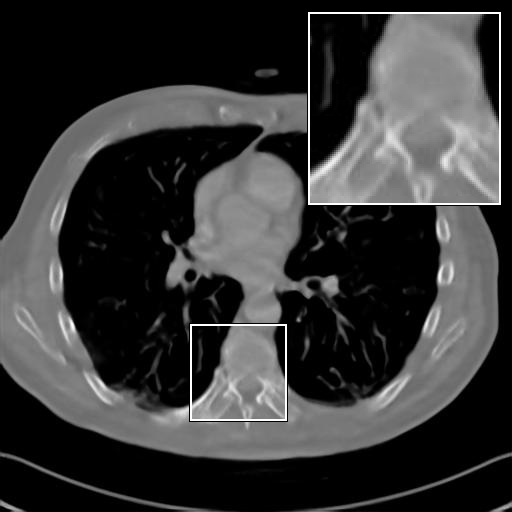}}
        \centerline{\scriptsize\shortstack{(k) Mod. BP + Ref. 1\\20.12/0.904}}\medskip
    \end{minipage}
    \begin{minipage}[b]{.19\linewidth}
        \centering
        \centerline{\includegraphics[width=\linewidth]{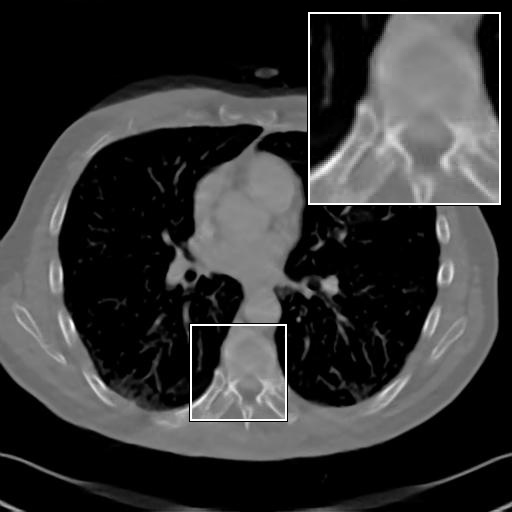}}
        \centerline{\scriptsize\shortstack{(l) Mod. BP + Ref. 2\\18.18/0.902}}\medskip
    \end{minipage}

    \begin{minipage}[b]{.19\linewidth}
        \centering
        \centerline{\includegraphics[width=\linewidth]{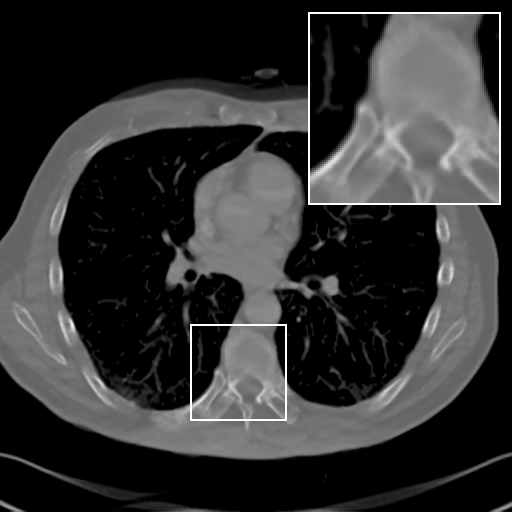}}
        \centerline{\scriptsize\shortstack{(m) PIE-DIP No Ref.\\17.98/0.910}}\medskip
    \end{minipage}
    \begin{minipage}[b]{.19\linewidth}
        \centering
        \centerline{\includegraphics[width=\linewidth]{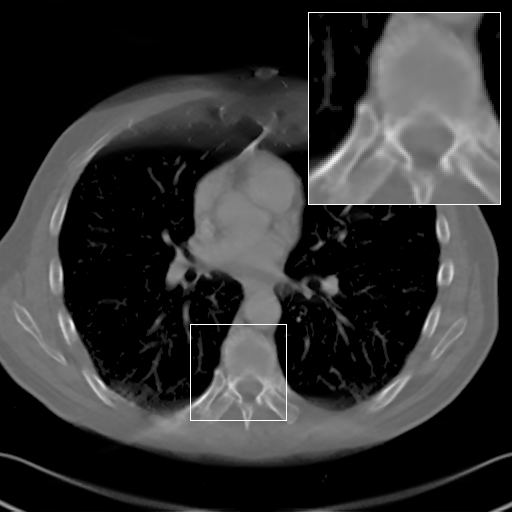}}
        \centerline{\scriptsize\shortstack{(n) PIE-DIP + ASD\\14.84/0.878}}\medskip
    \end{minipage}
    \begin{minipage}[b]{.19\linewidth}
        \centering
        \centerline{\includegraphics[width=\linewidth]{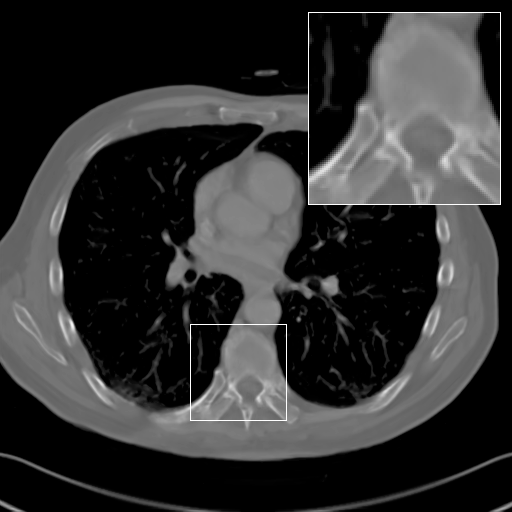}}
        \centerline{\scriptsize\shortstack{(o) PIE-DIP + Ref. 1\\22.38/0.954}}\medskip
    \end{minipage}
    \begin{minipage}[b]{.19\linewidth}
        \centering
        \centerline{\includegraphics[width=\linewidth]{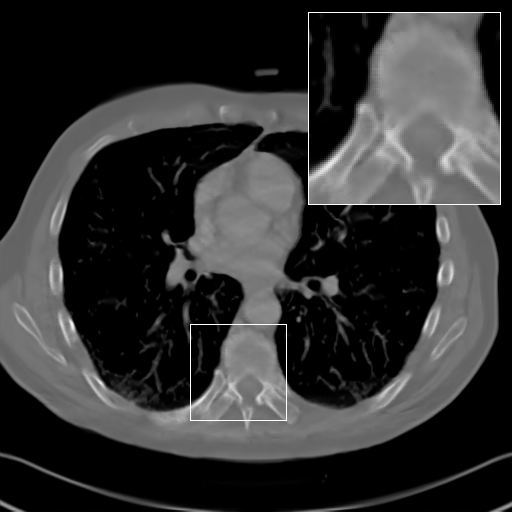}}
        \centerline{\scriptsize\shortstack{(p) PIE-DIP + Ref. 2\\20.24/0.920}}\medskip
    \end{minipage}

    \caption{Ablation study of the network input and patch-group prior in limited-angle CT reconstruction. The columns correspond to no reference image, an ASD-POCS reconstruction, Reference 1, and Reference 2. From top to bottom, the rows show the source images, direct input substitution using CDIP, the proposed modified back-projection input without the patch-group prior, and PIE-DIP with the patch-group prior. In columns 2--4, the corresponding reference image modulates the modified back-projection input and supplies reference patches to the mixed dictionary. In the first column, CDIP uses Gaussian noise and reduces to standard DIP; panel (i) is the modified-input-only baseline, whereas panel (m) additionally uses a self-only patch dictionary constructed from the evolving reconstruction with $P_i=0$. Values are reported as SNR (dB)/SSIM. The experiment uses 240 projections uniformly distributed over \(0^{\circ}\) to \(120^{\circ}\).}
    \label{netinput}
\end{figure*}

Fig.~\ref{netinput} compares no reference image, an artifact-contaminated ASD-POCS reconstruction, Reference 1, and the less similar Reference 2. Ground truth in panel (a) is used only for evaluation. Rows 2--4 show DIP/CDIP, modified-input DIP without the patch prior, and complete PIE-DIP, respectively. The no-reference DIP baseline uses Gaussian noise.

The modified input follows Eq.~\eqref{eq:modified_bp_input} without a reference and Eq.~\eqref{eq:modified_bp_input_ref} otherwise. In row 4, the no-reference case uses a self-only dictionary (\(P_i=0\)); in the other columns, the corresponding reference supplies both input modulation and dictionary patches.

The modified input increases SNR from 13.99 to 17.27~dB without a reference, from 13.45 to 14.14~dB with ASD-POCS, from 19.19 to 20.12~dB with Reference 1, and from 16.02 to 18.18~dB with Reference 2. CDIP with the artifact-contaminated ASD-POCS input gives lower SNR than standard DIP, illustrating sensitivity to reference quality. Severe artifacts in the modulating image also limit the benefit of the modified input.

Adding the patch prior increases SNR from 17.27 to 17.98~dB and SSIM from 0.879 to 0.910 without a reference. With ASD-POCS, Reference 1, and Reference 2, SNR rises from 14.14 to 14.84~dB, from 20.12 to 22.38~dB, and from 18.18 to 20.24~dB, respectively. Thus, self-only guidance is useful, while References 1 and 2 provide larger improvements.

\subsubsection{Effect of the Collaborative Patch-Group Prior}
\begin{figure*}[htbp]
    \centering
    \begin{minipage}[b]{.19\linewidth}
        \centering
        \includegraphics[width=\linewidth]{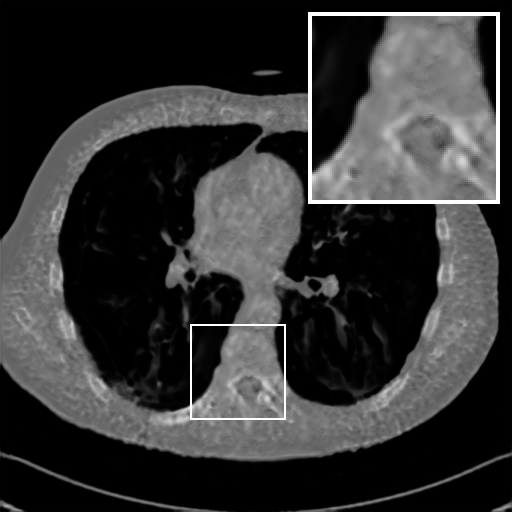}
        \centerline{\scriptsize\shortstack{(a) Modified-input DIP\\19.67/0.849}}\medskip
    \end{minipage}
    \begin{minipage}[b]{.19\linewidth}
        \centering
        \includegraphics[width=\linewidth]{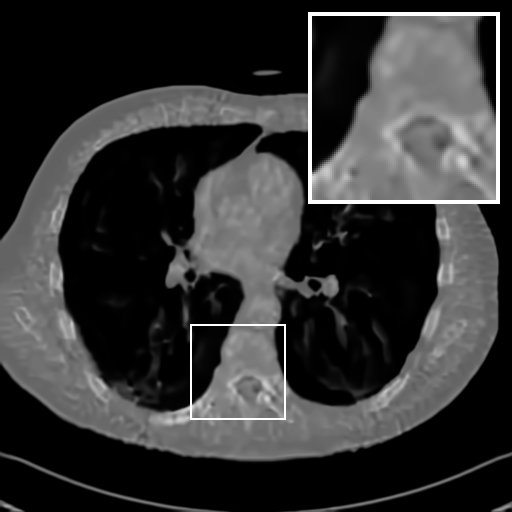}
        \centerline{\scriptsize\shortstack{(b) BM3D\\20.28/0.872}}\medskip
    \end{minipage}
    \begin{minipage}[b]{.19\linewidth}
        \centering
        \includegraphics[width=\linewidth]{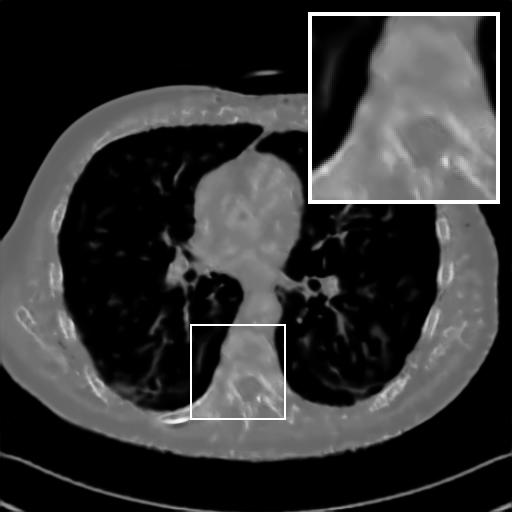}
        \centerline{\scriptsize\shortstack{(c) Self-only prior\\20.73/0.882}}\medskip
    \end{minipage}
    \begin{minipage}[b]{.19\linewidth}
        \centering
        \includegraphics[width=\linewidth]{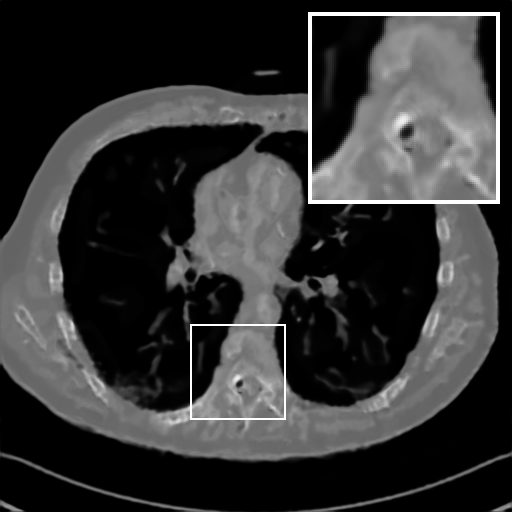}
        \centerline{\scriptsize\shortstack{(d) Reference-only prior\\20.74/0.891}}\medskip
    \end{minipage}
    \begin{minipage}[b]{.19\linewidth}
        \centering
        \includegraphics[width=\linewidth]{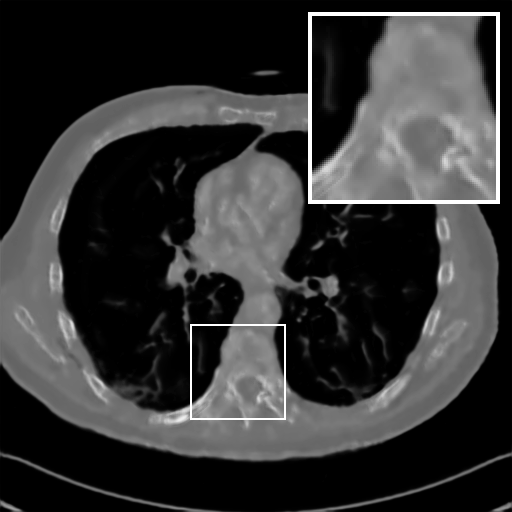}
        \centerline{\scriptsize\shortstack{(e) PIE-DIP\\22.26/0.925}}\medskip
    \end{minipage}

    \begin{minipage}[b]{.19\linewidth}
        \centering
        \includegraphics[width=\linewidth]{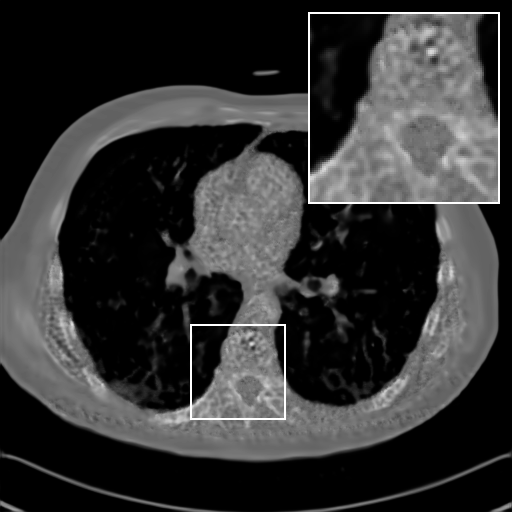}
        \centerline{\scriptsize\shortstack{(f) Modified-input DIP\\19.24/0.834}}\medskip
    \end{minipage}
    \begin{minipage}[b]{.19\linewidth}
        \centering
        \includegraphics[width=\linewidth]{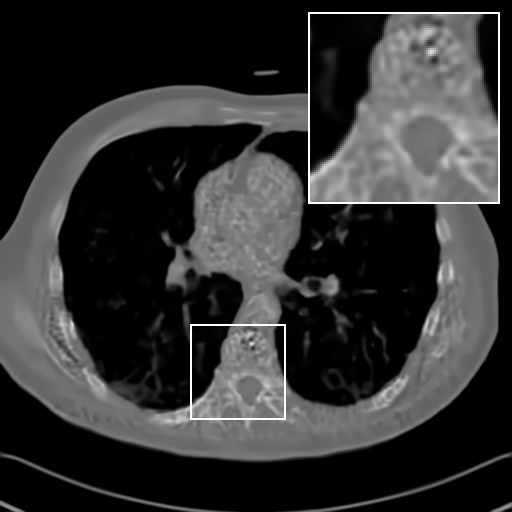}
        \centerline{\scriptsize\shortstack{(g) BM3D\\20.15/0.865}}\medskip
    \end{minipage}
    \begin{minipage}[b]{.19\linewidth}
        \centering
        \includegraphics[width=\linewidth]{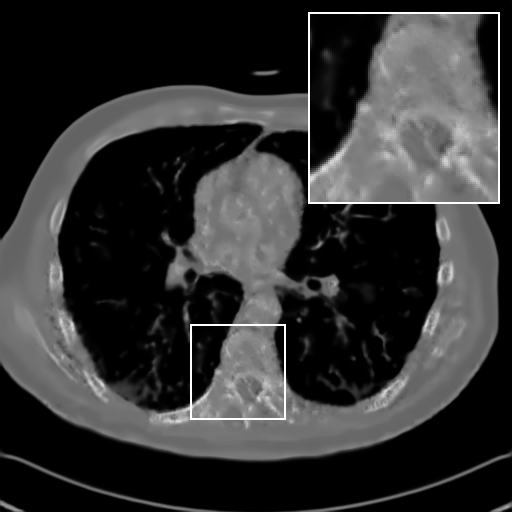}
        \centerline{\scriptsize\shortstack{(h) Self-only prior\\20.97/0.693}}\medskip
    \end{minipage}
    \begin{minipage}[b]{.19\linewidth}
        \centering
        \includegraphics[width=\linewidth]{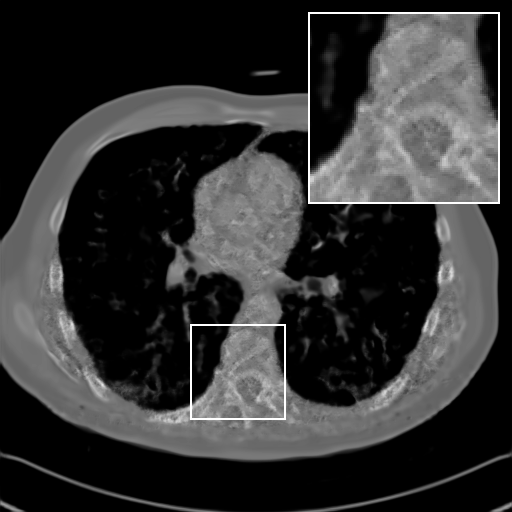}
        \centerline{\scriptsize\shortstack{(i) Reference-only prior\\20.90/0.869}}\medskip
    \end{minipage}
    \begin{minipage}[b]{.19\linewidth}
        \centering
        \includegraphics[width=\linewidth]{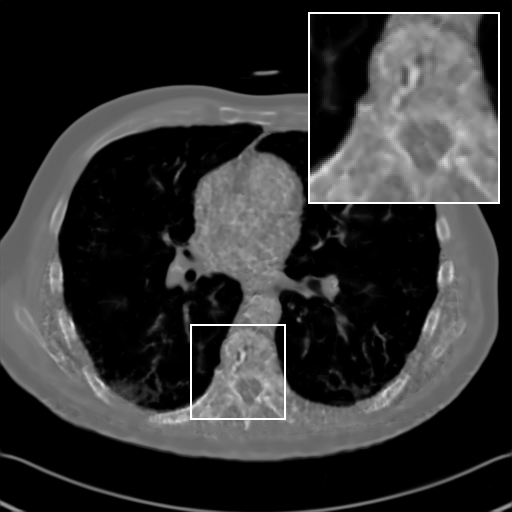}
        \centerline{\scriptsize\shortstack{(j) PIE-DIP\\21.61/0.903}}\medskip
    \end{minipage}

    \caption{Ablation study of the collaborative patch-group prior for sparse-view CT reconstruction of the target in Fig.~\ref{netinput}(a). The first and second rows use the images in Fig.~\ref{netinput}(c) and Fig.~\ref{netinput}(d) as Reference 1 and Reference 2, respectively. From left to right, the columns show modified-input DIP without patch processing, BM3D, the proposed patch-group prior with a self-only dictionary (\(P_i=0\)), the patch-group prior with a reference-only dictionary (\(P_i=1\)), and complete PIE-DIP with the scheduled mixed dictionary. Values are reported as SNR (dB)/SSIM. Magnified regions highlight the preservation of local anatomical details. The experiment uses 30 projections uniformly distributed over \(0^\circ\) to \(360^\circ\).}
    \label{patchprior_ablation}
\end{figure*}

Fig.~\ref{patchprior_ablation} compares five configurations for the target in Fig.~\ref{netinput}(a), using 30 projections uniformly distributed over \(0^\circ\)--\(360^\circ\). The two rows use Reference 1 in Fig.~\ref{netinput}(c) and Reference 2 in Fig.~\ref{netinput}(d), respectively.

Modified-input DIP serves as the baseline for BM3D and patch guidance with self-only (\(P_i=0\)), reference-only (\(P_i=1\)), or scheduled mixed dictionaries. The DIP input remains fixed within each row, including reference modulation in the self-only case. BM3D is applied to the baseline reconstruction, with its strength selected on the validation subset.

BM3D increases SNR by 0.61 and 0.91~dB for References 1 and 2, respectively. Self-only guidance gives larger gains of 1.06 and approximately 1.73~dB, but SSIM rises from 0.849 to 0.882 in the first row and falls from 0.834 to 0.693 in the second. Self-only guidance therefore does not consistently improve SSIM and may reinforce early reconstruction artifacts.

Reference-only guidance increases SNR by 1.07 and 1.66~dB in Fig.~\ref{patchprior_ablation}(d) and (i). Gains with the less similar Reference 2 show that local matching can extract useful patches despite global differences, although exclusive reliance on reference patches may limit later refinement.

The scheduled mixed dictionary combines both sources and achieves the highest SNR and SSIM in both rows. SNR rises from 19.67 to 22.26~dB with Reference 1 and from 19.24 to 21.61~dB with Reference 2, exceeding the reference-only results by 1.52 and 0.71~dB, respectively.

\subsubsection{Optimization Behavior}
\begin{figure}[htbp]
    \centering
    \begin{minipage}[b]{0.48\linewidth}
        \centering
        \centerline{\includegraphics[width=\linewidth]{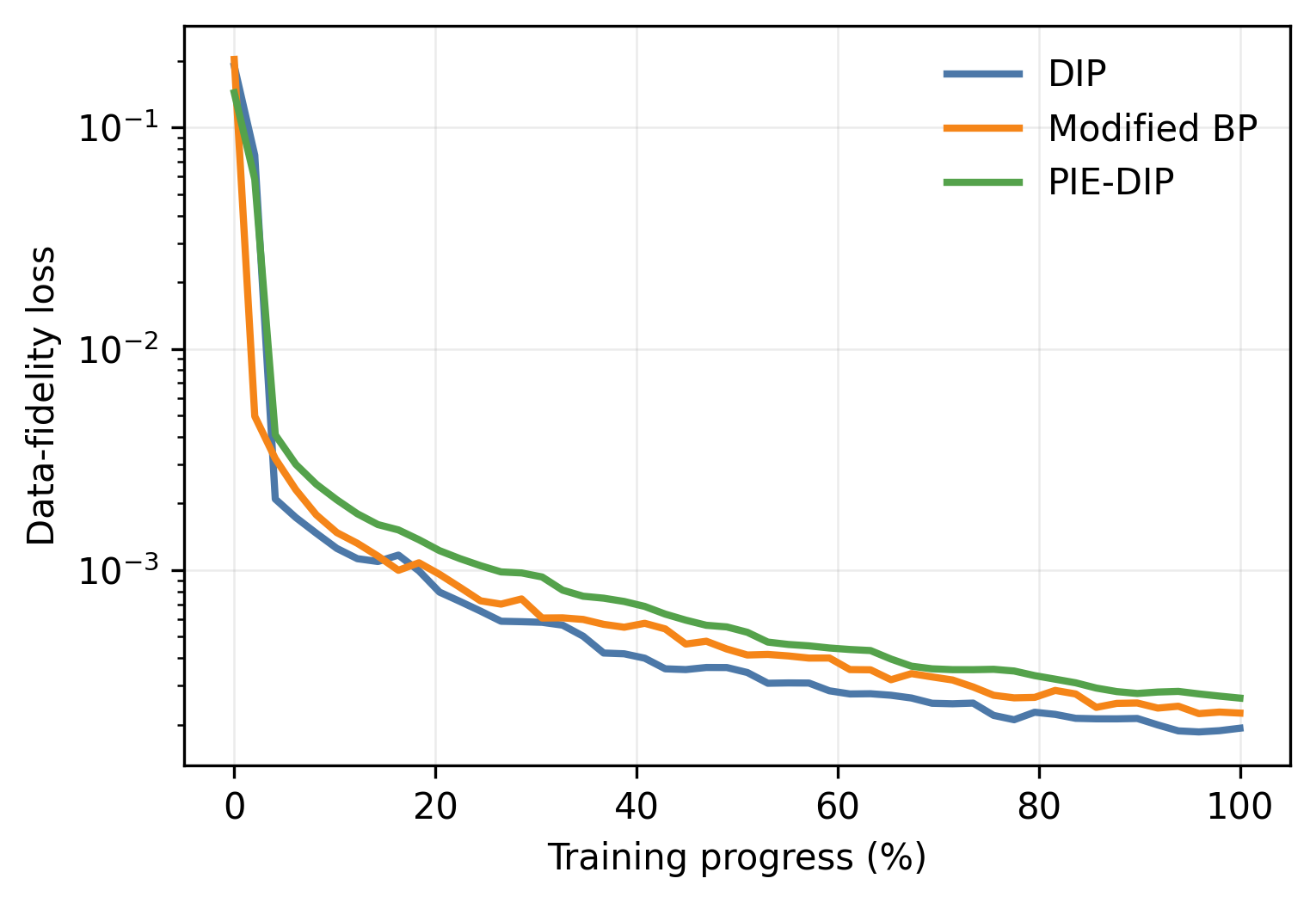}}
        \centerline{(a)}\medskip
    \end{minipage}\hfill
    \begin{minipage}[b]{0.48\linewidth}
        \centering
        \centerline{\includegraphics[width=\linewidth]{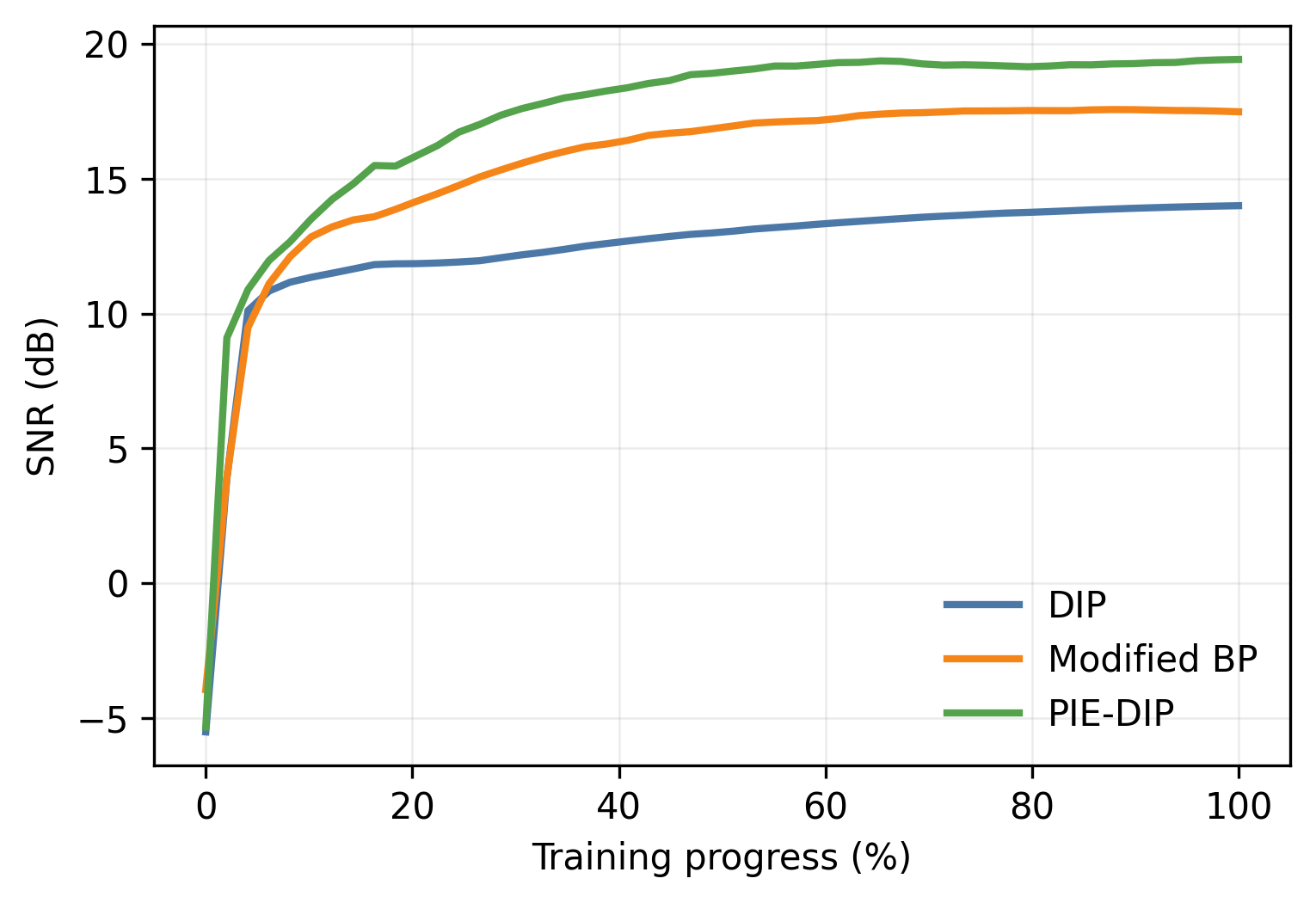}}
        \centerline{(b)}\medskip
    \end{minipage}
    \caption{(a) Mean data-fidelity loss on a logarithmic scale and (b) mean SNR for limited-angle CT reconstruction. The curves compare standard DIP, the proposed modified back-projection input, and complete PIE-DIP with the patch-group prior.}
    \label{trainingperformance}
\end{figure}

Fig.~\ref{trainingperformance} compares reconstruction SNR and the data-fidelity loss, excluding the patch-group term in Eq.~\eqref{eq:piedip_objective}. Standard DIP achieves the lowest loss, followed by modified-input DIP and complete PIE-DIP, whereas SNR follows the reverse ordering. This reversal demonstrates that a smaller projection residual does not, by itself, identify a more accurate reconstruction under the tested limited-angle conditions.

Limited angular coverage leaves some image components weakly constrained, so closer agreement with the measured projections can still accompany substantial reconstruction error. The modified input and patch-group prior provide complementary structural guidance, each improving SNR despite a larger data-fidelity residual. These gains demonstrate the added value of the proposed prior design beyond projection fitting alone.

The PIE-DIP curves also show an overall decrease in data-fidelity loss and a stable SNR plateau in later iterations, providing empirical evidence of convergence in reconstruction accuracy under the tested conditions.

\subsection{Sparse-View and Limited-Angle CT Reconstruction}\label{sec:LAFWRecon}
Experiments use 100 patient cases randomly sampled directly from LIDC-IDRI~\cite{armato2011lung}. Sparse-view reconstruction uses 30, 60, 90, 120, 180, and 360 projections uniformly distributed over \(0^\circ\)--\(360^\circ\); limited-angle reconstruction covers \(90^\circ\)--\(180^\circ\) with two projections per degree.

ASD-POCS, DIP, and CDIP provide baselines for conventional iterative reconstruction, the implicit network prior, and direct reference input, respectively. PICCS~\cite{chen2008piccs} and NCPICCS~\cite{ramirez2011ncpiccs} represent global reference-difference constraints. CvG-Diff for sparse-view CT~\cite{chen2025cvgdiff} and DOLCE for limited-angle CT~\cite{liu2023dolce} provide pretrained diffusion baselines with learned image priors; each is trained separately using images from 500 patients.

All methods use the same projection data. CDIP, PIE-DIP, PICCS, and NCPICCS also share a reference CT slice selected from the database for each case. PICCS combines image and reference-difference sparsity, whereas NCPICCS uses nonconvex \(\ell_p\) penalties. Their regularization parameters are selected on a separate validation subset and fixed for testing.

\begin{figure*}
    \centering
    \begin{minipage}[b]{.104\linewidth}
        \centering
        \centerline{\includegraphics[width=\linewidth]{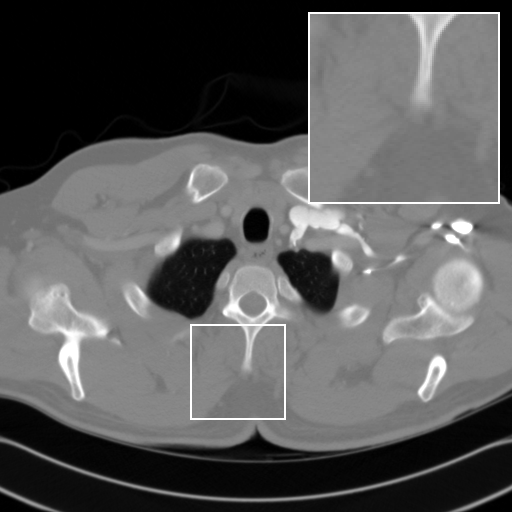}}
        \centerline{\tiny Ground truth}
        \centerline{\tiny --/--}\medskip
    \end{minipage}
    \begin{minipage}[b]{.104\linewidth}
        \centering
        \centerline{\includegraphics[width=\linewidth]{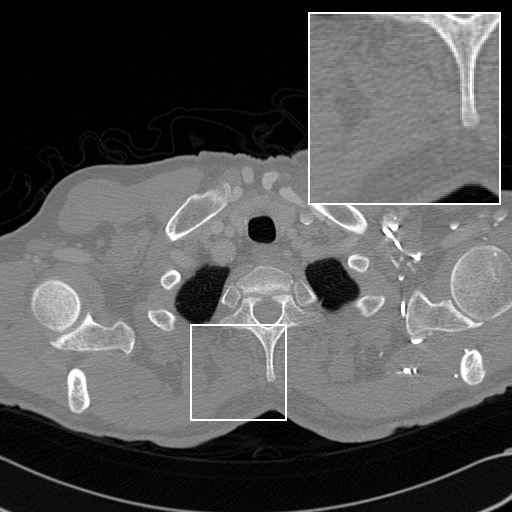}}
        \centerline{\tiny Reference}
        \centerline{\tiny --/--}\medskip
    \end{minipage}
    \begin{minipage}[b]{.104\linewidth}
        \centering
        \centerline{\includegraphics[width=\linewidth]{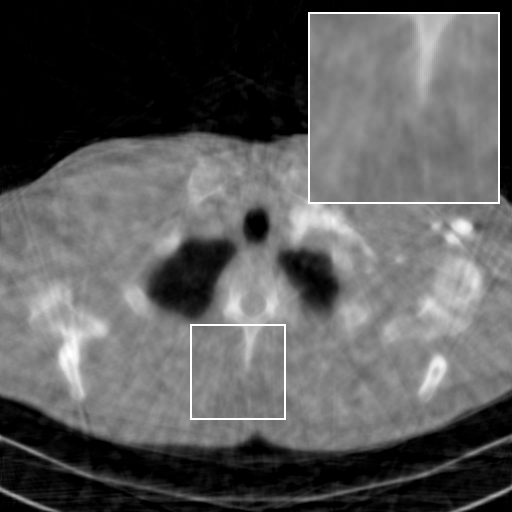}}
        \centerline{\tiny ASD-POCS}
        \centerline{\tiny 19.59/0.761}\medskip
    \end{minipage}
    \begin{minipage}[b]{.104\linewidth}
        \centering
        \centerline{\includegraphics[width=\linewidth]{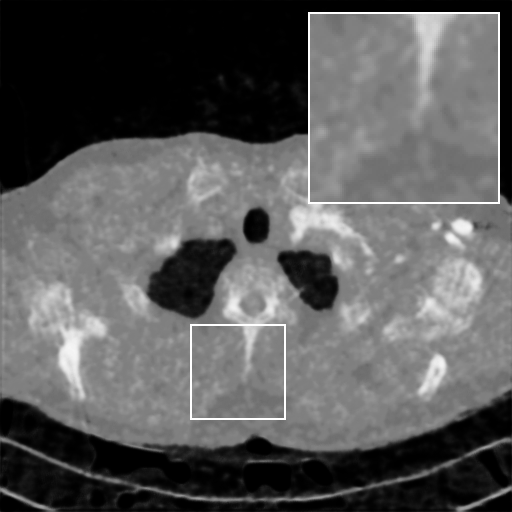}}
        \centerline{\tiny DIP}
        \centerline{\tiny 21.27/0.800}\medskip
    \end{minipage}
    \begin{minipage}[b]{.104\linewidth}
        \centering
        \centerline{\includegraphics[width=\linewidth]{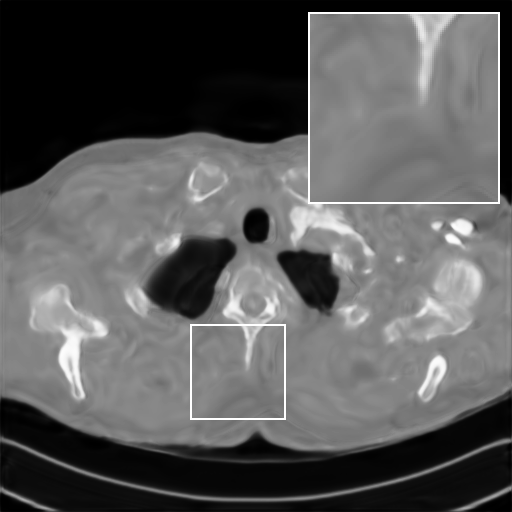}}
        \centerline{\tiny CDIP}
        \centerline{\tiny 25.18/0.897}\medskip
    \end{minipage}
    \begin{minipage}[b]{.104\linewidth}
        \centering
        \centerline{\includegraphics[width=\linewidth]{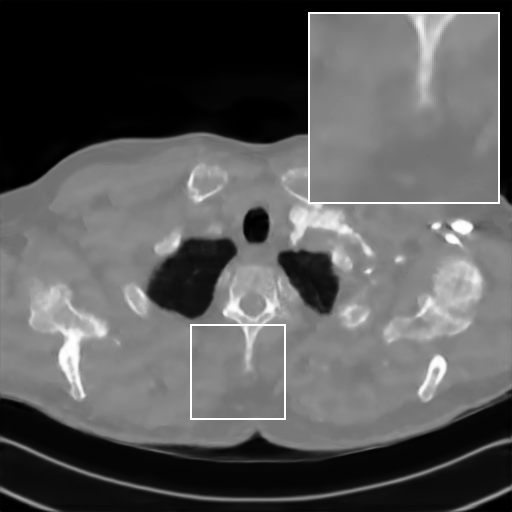}}
        \centerline{\tiny PIE-DIP}
        \centerline{\tiny 26.67/0.922}\medskip
    \end{minipage}
    \begin{minipage}[b]{.104\linewidth}
        \centering
        \centerline{\includegraphics[width=\linewidth]{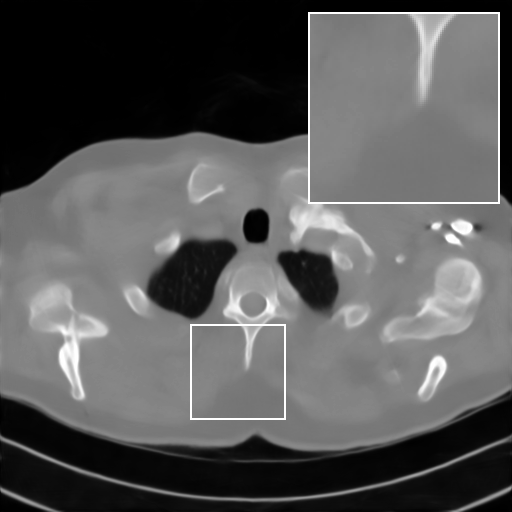}}
        \centerline{\tiny CvG-Diff}
        \centerline{\tiny 24.38/0.906}\medskip
    \end{minipage}
    \begin{minipage}[b]{.104\linewidth}
        \centering
        \centerline{\includegraphics[width=\linewidth]{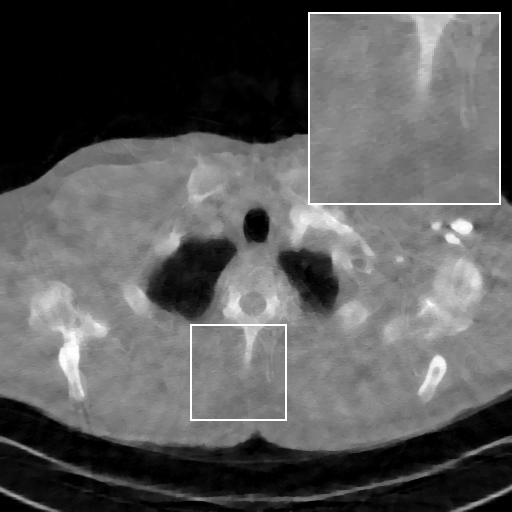}}
        \centerline{\tiny PICCS}
        \centerline{\tiny 21.02/0.809}\medskip
    \end{minipage}
    \begin{minipage}[b]{.104\linewidth}
        \centering
        \centerline{\includegraphics[width=\linewidth]{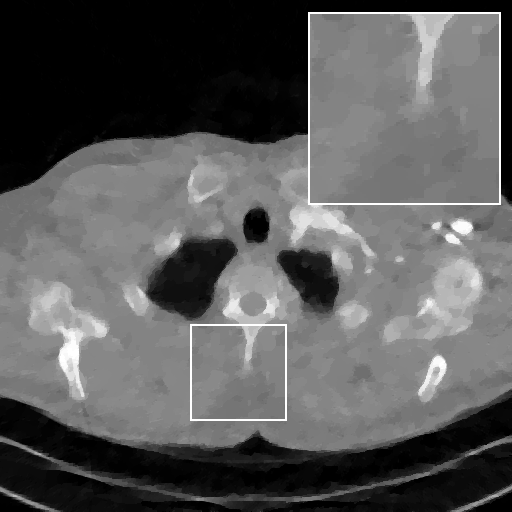}}
        \centerline{\tiny NCPICCS}
        \centerline{\tiny 21.51/0.844}\medskip
    \end{minipage}
    \par\smallskip
    \begin{minipage}[b]{.104\linewidth}
        \centering
        \centerline{\includegraphics[width=\linewidth]{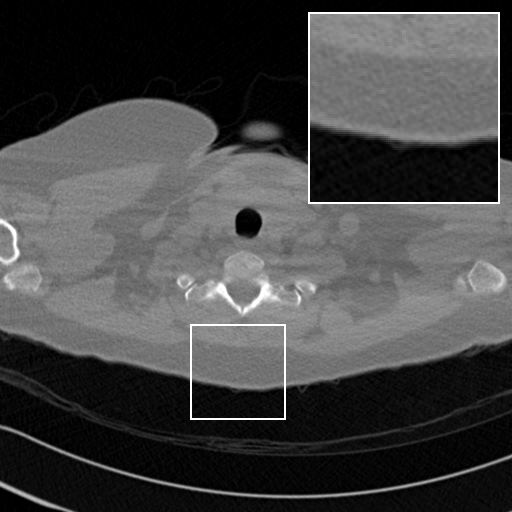}}
        \centerline{\tiny Ground truth}
        \centerline{\tiny --/--}\medskip
    \end{minipage}
    \begin{minipage}[b]{.104\linewidth}
        \centering
        \centerline{\includegraphics[width=\linewidth]{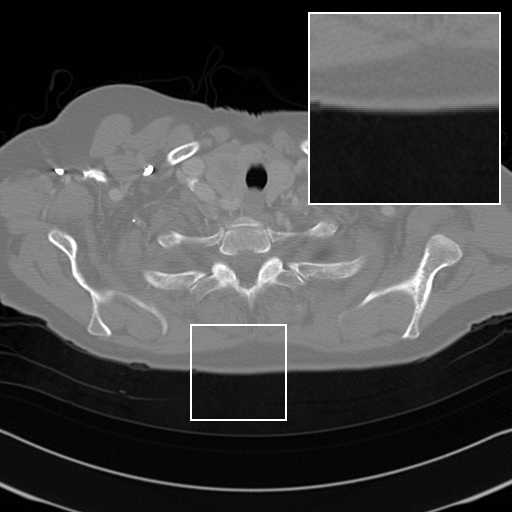}}
        \centerline{\tiny Reference}
        \centerline{\tiny --/--}\medskip
    \end{minipage}
    \begin{minipage}[b]{.104\linewidth}
        \centering
        \centerline{\includegraphics[width=\linewidth]{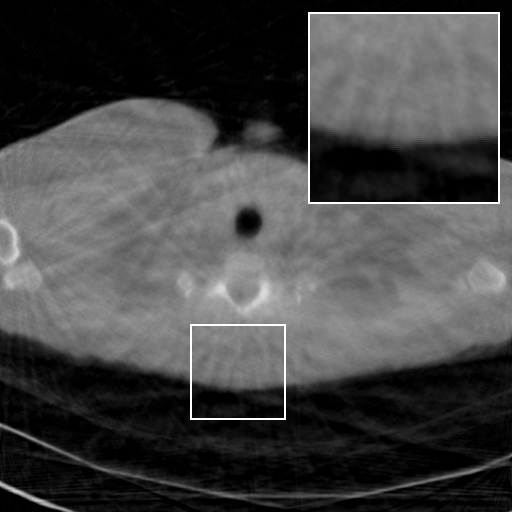}}
        \centerline{\tiny ASD-POCS}
        \centerline{\tiny 17.89/0.692}\medskip
    \end{minipage}
    \begin{minipage}[b]{.104\linewidth}
        \centering
        \centerline{\includegraphics[width=\linewidth]{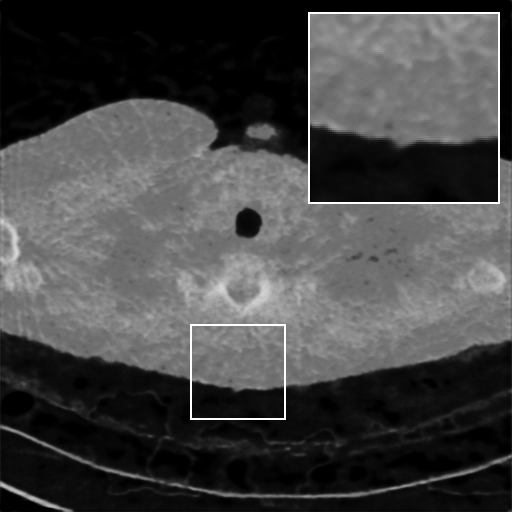}}
        \centerline{\tiny DIP}
        \centerline{\tiny 19.51/0.700}\medskip
    \end{minipage}
    \begin{minipage}[b]{.104\linewidth}
        \centering
        \centerline{\includegraphics[width=\linewidth]{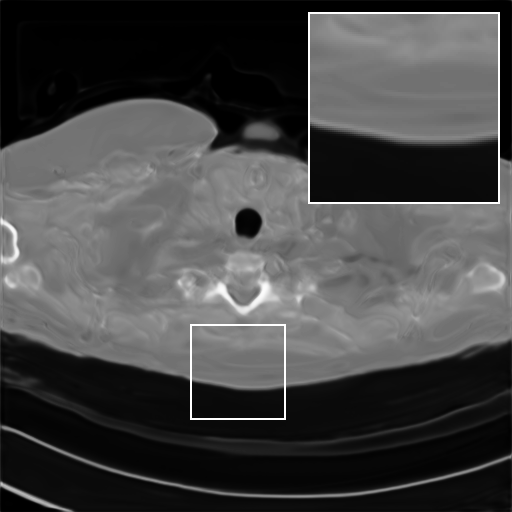}}
        \centerline{\tiny CDIP}
        \centerline{\tiny 22.33/0.784}\medskip
    \end{minipage}
    \begin{minipage}[b]{.104\linewidth}
        \centering
        \centerline{\includegraphics[width=\linewidth]{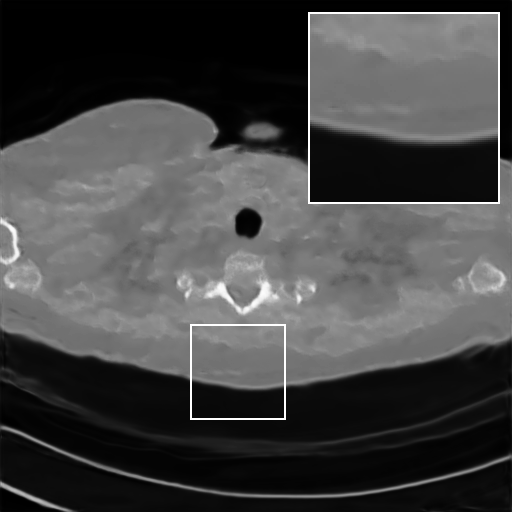}}
        \centerline{\tiny PIE-DIP}
        \centerline{\tiny 22.99/0.808}\medskip
    \end{minipage}
    \begin{minipage}[b]{.104\linewidth}
        \centering
        \centerline{\includegraphics[width=\linewidth]{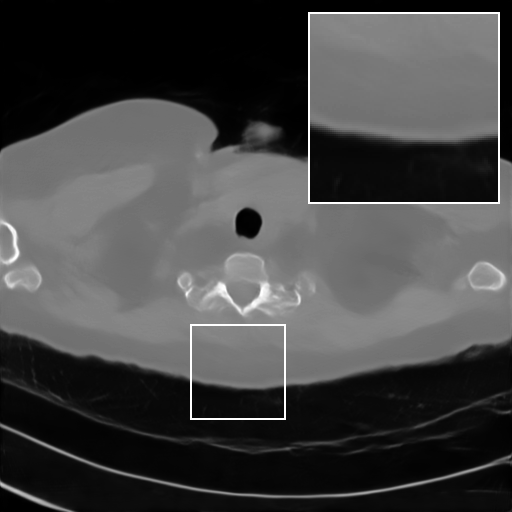}}
        \centerline{\tiny CvG-Diff}
        \centerline{\tiny 21.51/0.793}\medskip
    \end{minipage}
    \begin{minipage}[b]{.104\linewidth}
        \centering
        \centerline{\includegraphics[width=\linewidth]{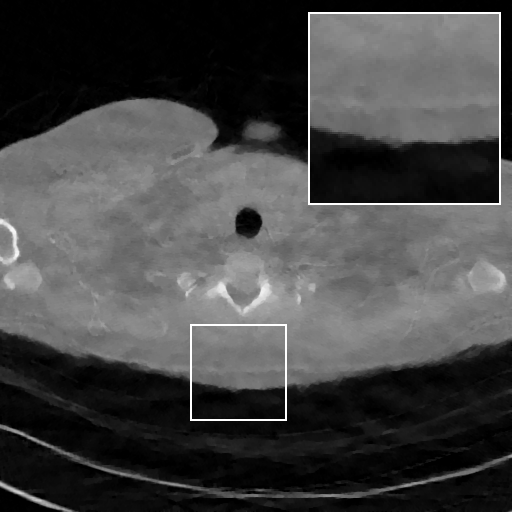}}
        \centerline{\tiny PICCS}
        \centerline{\tiny 19.78/0.761}\medskip
    \end{minipage}
    \begin{minipage}[b]{.104\linewidth}
        \centering
        \centerline{\includegraphics[width=\linewidth]{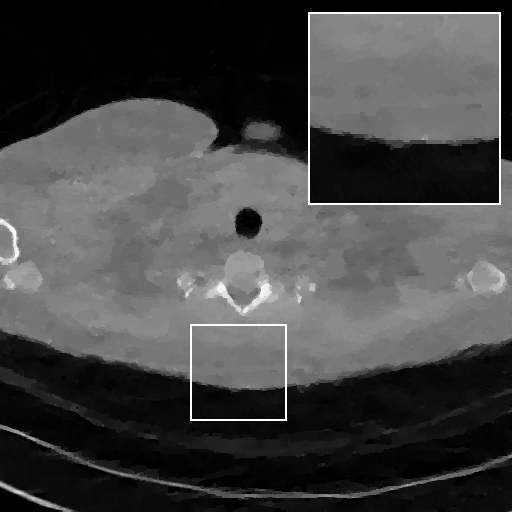}}
        \centerline{\tiny NCPICCS}
        \centerline{\tiny 19.76/0.772}\medskip
    \end{minipage}
    \par\smallskip
    \begin{minipage}[b]{.104\linewidth}
        \centering
        \centerline{\includegraphics[width=\linewidth]{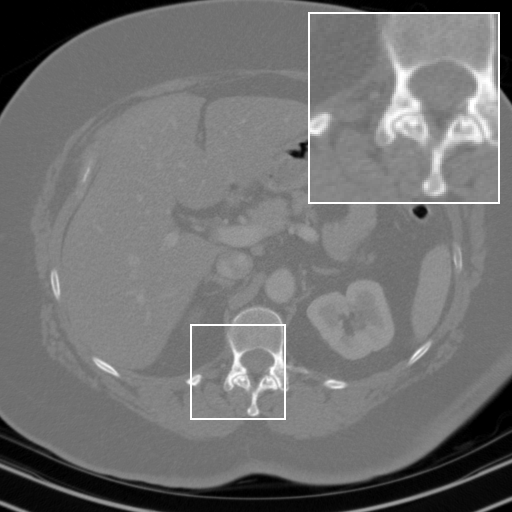}}
        \centerline{\tiny Ground truth}
        \centerline{\tiny --/--}\medskip
    \end{minipage}
    \begin{minipage}[b]{.104\linewidth}
        \centering
        \centerline{\includegraphics[width=\linewidth]{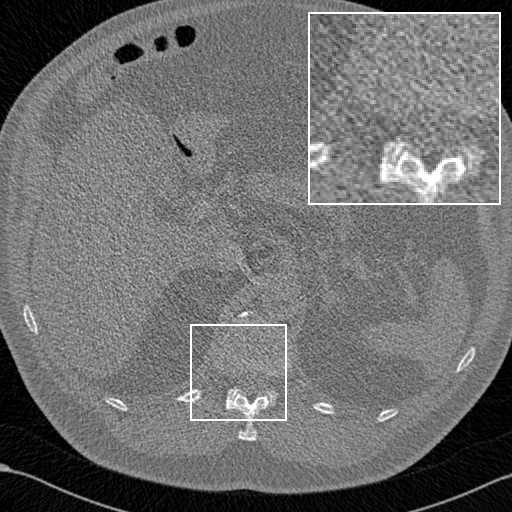}}
        \centerline{\tiny Reference}
        \centerline{\tiny --/--}\medskip
    \end{minipage}
    \begin{minipage}[b]{.104\linewidth}
        \centering
        \centerline{\includegraphics[width=\linewidth]{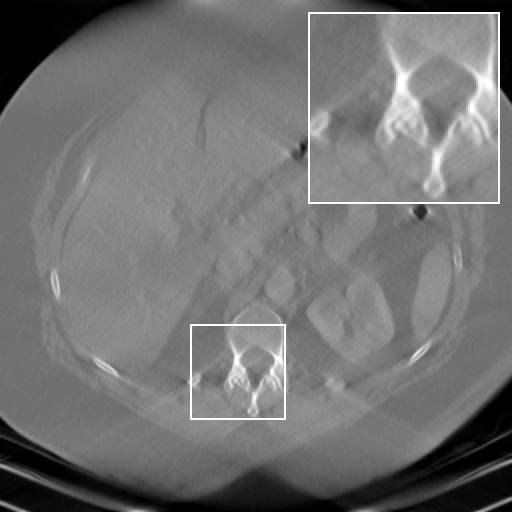}}
        \centerline{\tiny ASD-POCS}
        \centerline{\tiny 20.30/0.875}\medskip
    \end{minipage}
    \begin{minipage}[b]{.104\linewidth}
        \centering
        \centerline{\includegraphics[width=\linewidth]{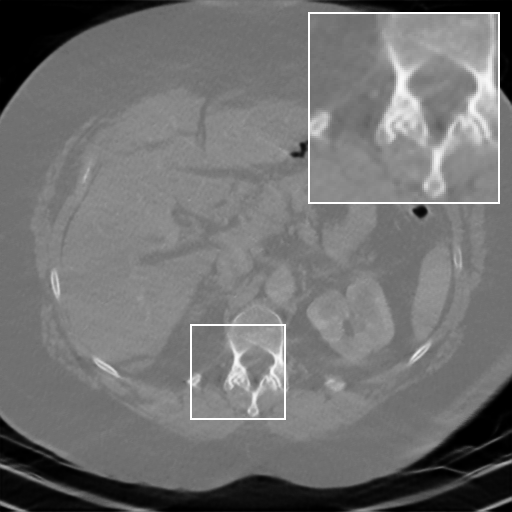}}
        \centerline{\tiny DIP}
        \centerline{\tiny 23.37/0.873}\medskip
    \end{minipage}
    \begin{minipage}[b]{.104\linewidth}
        \centering
        \centerline{\includegraphics[width=\linewidth]{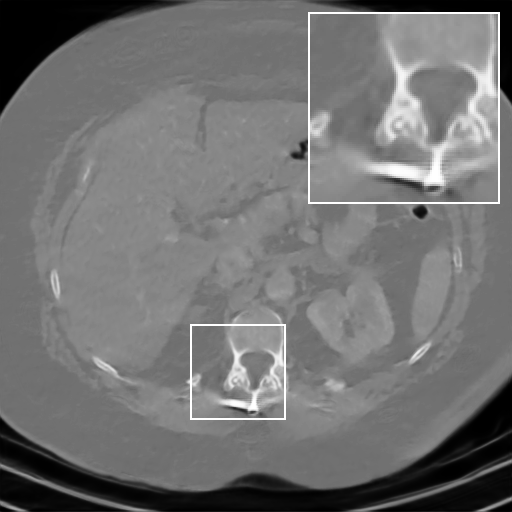}}
        \centerline{\tiny CDIP}
        \centerline{\tiny 22.95/0.866}\medskip
    \end{minipage}
    \begin{minipage}[b]{.104\linewidth}
        \centering
        \centerline{\includegraphics[width=\linewidth]{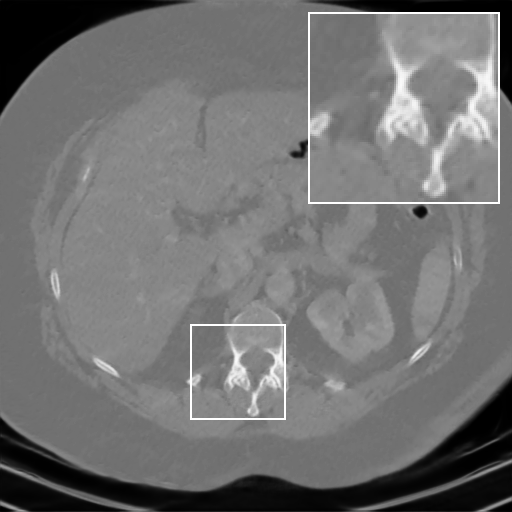}}
        \centerline{\tiny PIE-DIP}
        \centerline{\tiny 24.45/0.890}\medskip
    \end{minipage}
    \begin{minipage}[b]{.104\linewidth}
        \centering
        \centerline{\includegraphics[width=\linewidth]{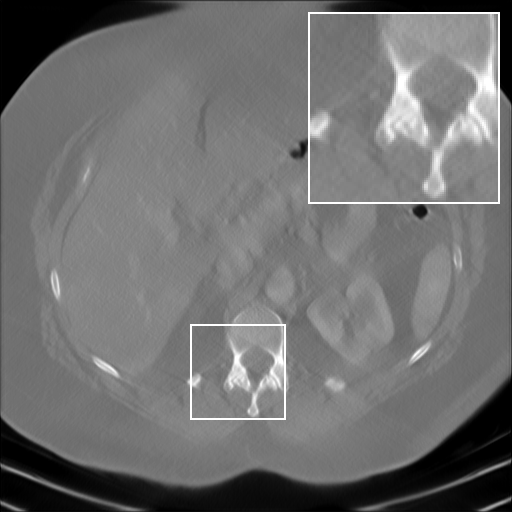}}
        \centerline{\tiny DOLCE}
        \centerline{\tiny 14.64/0.767}\medskip
    \end{minipage}
    \begin{minipage}[b]{.104\linewidth}
        \centering
        \centerline{\includegraphics[width=\linewidth]{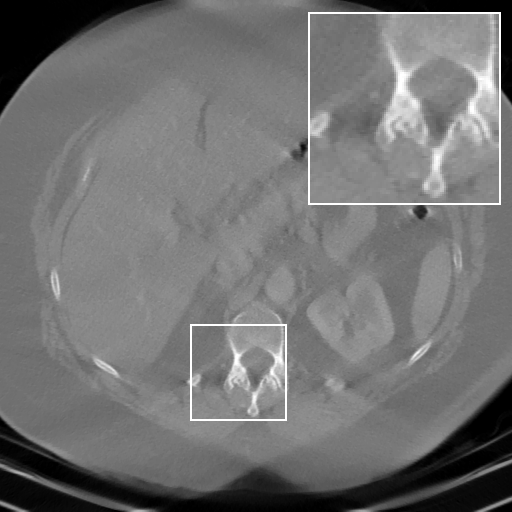}}
        \centerline{\tiny PICCS}
        \centerline{\tiny 21.41/0.870}\medskip
    \end{minipage}
    \begin{minipage}[b]{.104\linewidth}
        \centering
        \centerline{\includegraphics[width=\linewidth]{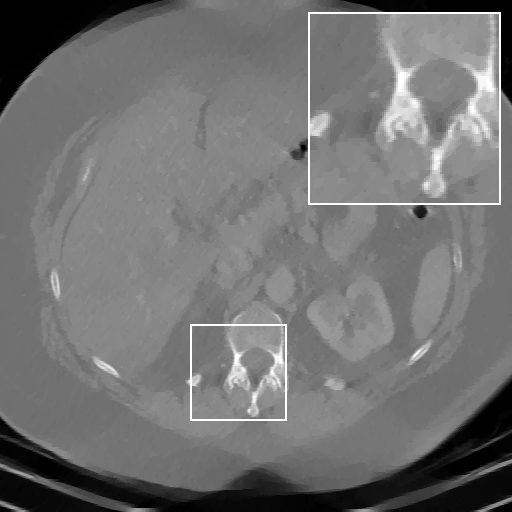}}
        \centerline{\tiny NCPICCS}
        \centerline{\tiny 21.58/0.871}\medskip
    \end{minipage}
    \par\smallskip
    \begin{minipage}[b]{.104\linewidth}
        \centering
        \centerline{\includegraphics[width=\linewidth]{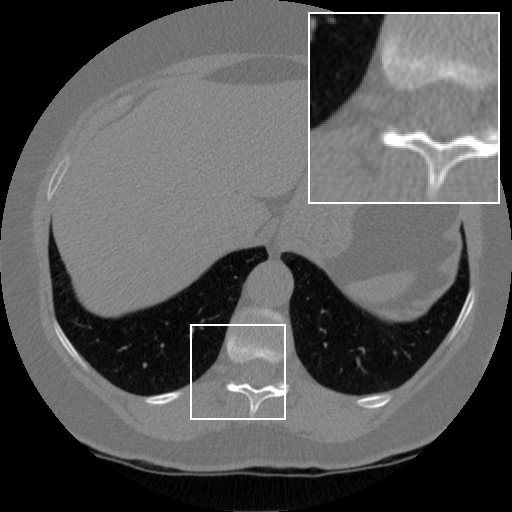}}
        \centerline{\tiny Ground truth}
        \centerline{\tiny --/--}\medskip
    \end{minipage}
    \begin{minipage}[b]{.104\linewidth}
        \centering
        \centerline{\includegraphics[width=\linewidth]{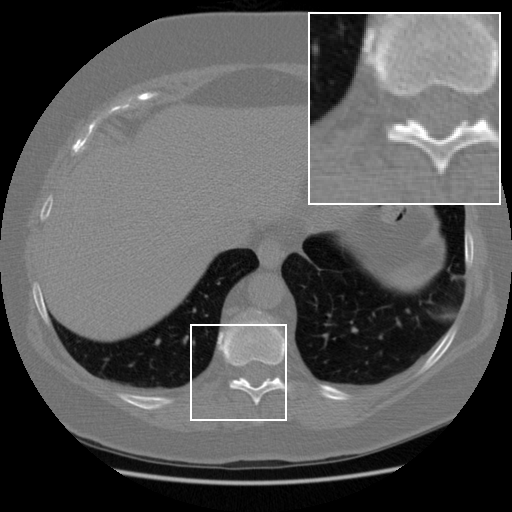}}
        \centerline{\tiny Reference}
        \centerline{\tiny --/--}\medskip
    \end{minipage}
    \begin{minipage}[b]{.104\linewidth}
        \centering
        \centerline{\includegraphics[width=\linewidth]{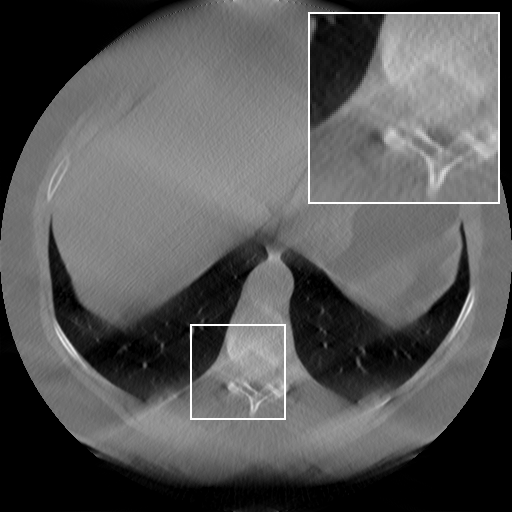}}
        \centerline{\tiny ASD-POCS}
        \centerline{\tiny 17.87/0.859}\medskip
    \end{minipage}
    \begin{minipage}[b]{.104\linewidth}
        \centering
        \centerline{\includegraphics[width=\linewidth]{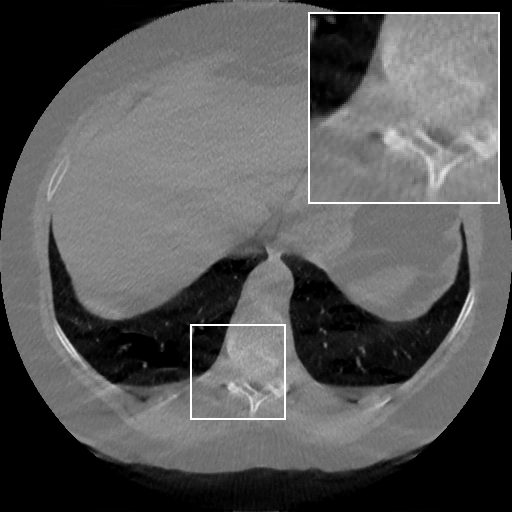}}
        \centerline{\tiny DIP}
        \centerline{\tiny 19.60/0.837}\medskip
    \end{minipage}
    \begin{minipage}[b]{.104\linewidth}
        \centering
        \centerline{\includegraphics[width=\linewidth]{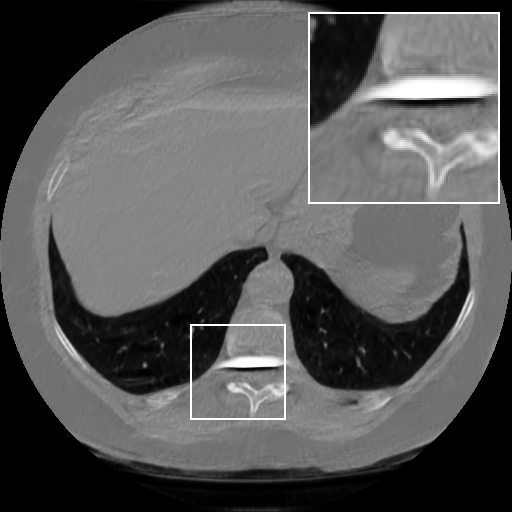}}
        \centerline{\tiny CDIP}
        \centerline{\tiny 20.03/0.829}\medskip
    \end{minipage}
    \begin{minipage}[b]{.104\linewidth}
        \centering
        \centerline{\includegraphics[width=\linewidth]{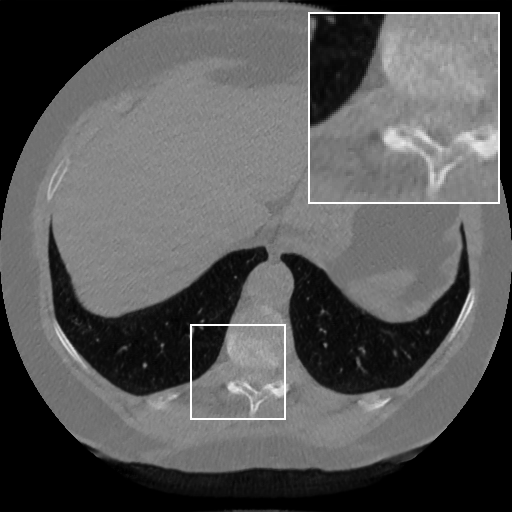}}
        \centerline{\tiny PIE-DIP}
        \centerline{\tiny 22.47/0.872}\medskip
    \end{minipage}
    \begin{minipage}[b]{.104\linewidth}
        \centering
        \centerline{\includegraphics[width=\linewidth]{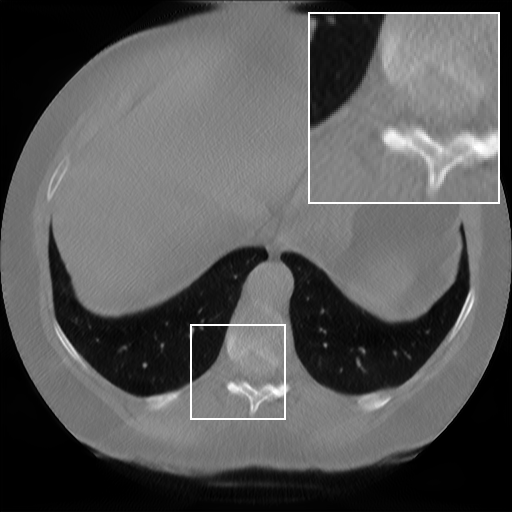}}
        \centerline{\tiny DOLCE}
        \centerline{\tiny 11.13/0.621}\medskip
    \end{minipage}
    \begin{minipage}[b]{.104\linewidth}
        \centering
        \centerline{\includegraphics[width=\linewidth]{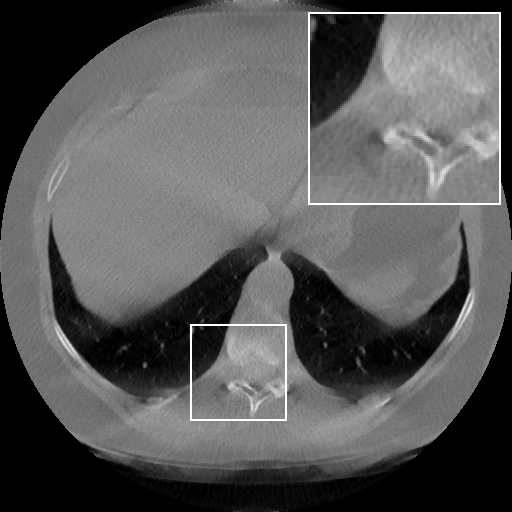}}
        \centerline{\tiny PICCS}
        \centerline{\tiny 19.40/0.871}\medskip
    \end{minipage}
    \begin{minipage}[b]{.104\linewidth}
        \centering
        \centerline{\includegraphics[width=\linewidth]{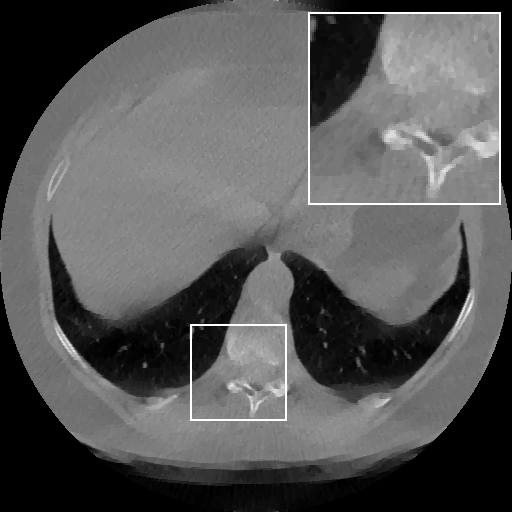}}
        \centerline{\tiny NCPICCS}
        \centerline{\tiny 19.74/0.845}\medskip
    \end{minipage}

    \caption{Representative LIDC-IDRI reconstruction results for two sparse-view cases and two limited-angle cases. The first two rows show sparse-view CT with 30 projections over \(0^{\circ}\) to \(360^{\circ}\), and the last two rows show limited-angle CT with a \(120^{\circ}\) angular range. From left to right, the columns show the ground truth, reference image, ASD-POCS, standard DIP, CDIP, PIE-DIP, the corresponding task-specific pretrained model (CvG-Diff for sparse-view CT and DOLCE for limited-angle CT), PICCS, and NCPICCS. Values below each reconstruction are reported as SNR (dB)/SSIM.}
    \label{reconresult}
\end{figure*}

Fig.~\ref{reconresult} shows two sparse-view and two limited-angle examples. PIE-DIP reduces undersampling artifacts and preserves anatomical detail more clearly than the iterative and DIP-based baselines in these cases. PICCS and NCPICCS remain affected by global reference differences, while PIE-DIP uses local patch matches. In these examples, CvG-Diff oversmooths sparse-view structures and DOLCE retains missing-angle artifacts in the limited-angle results; PIE-DIP achieves higher SNR without offline training.

\begin{figure}[htb]
    \centering
    \begin{minipage}[b]{0.48\linewidth}
        \centering
        \centerline{\includegraphics[width=\linewidth]{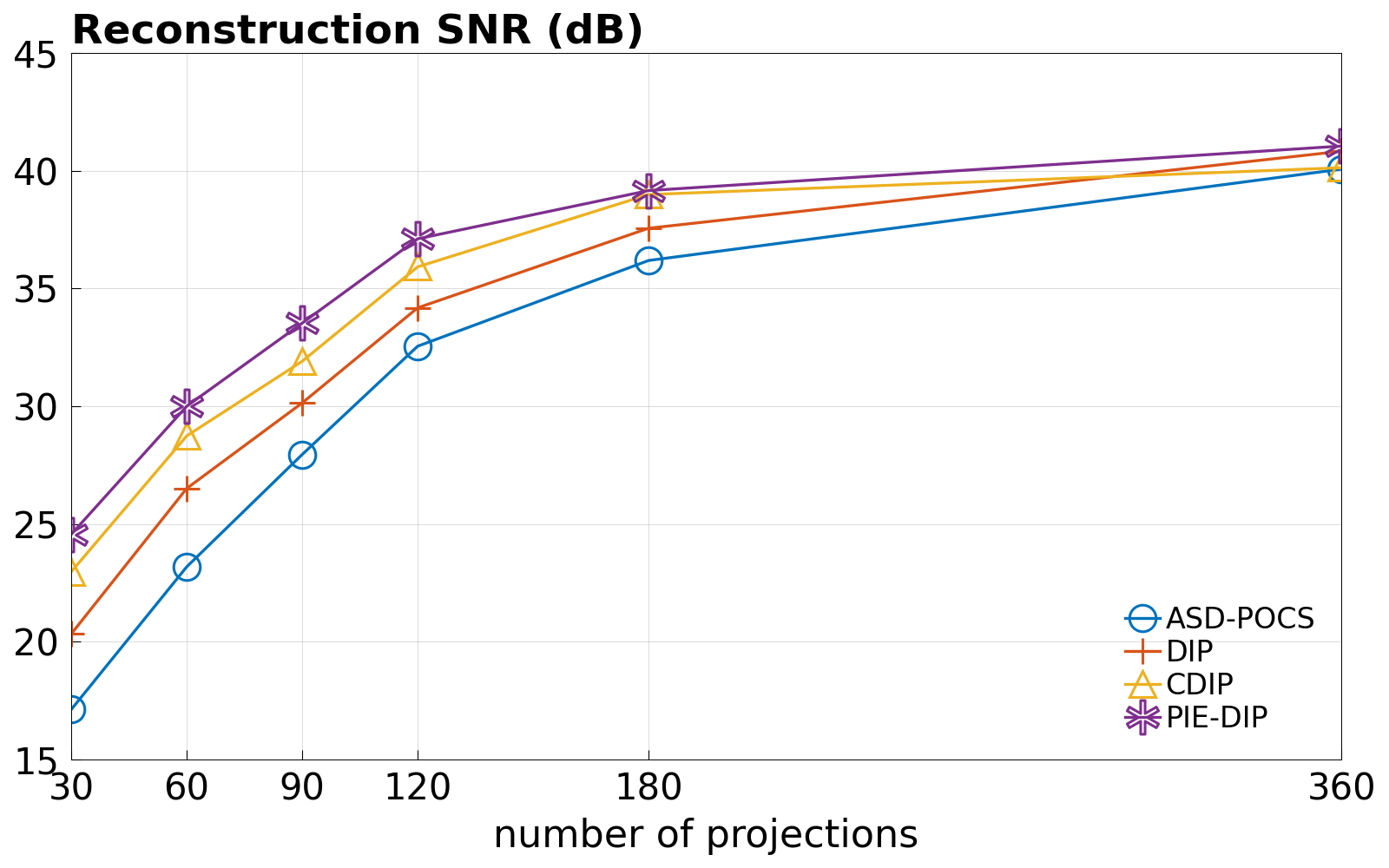}}
        \centerline{(a)}\medskip
    \end{minipage}\hfill%
    \begin{minipage}[b]{0.48\linewidth}
        \centering
        \centerline{\includegraphics[width=\linewidth]{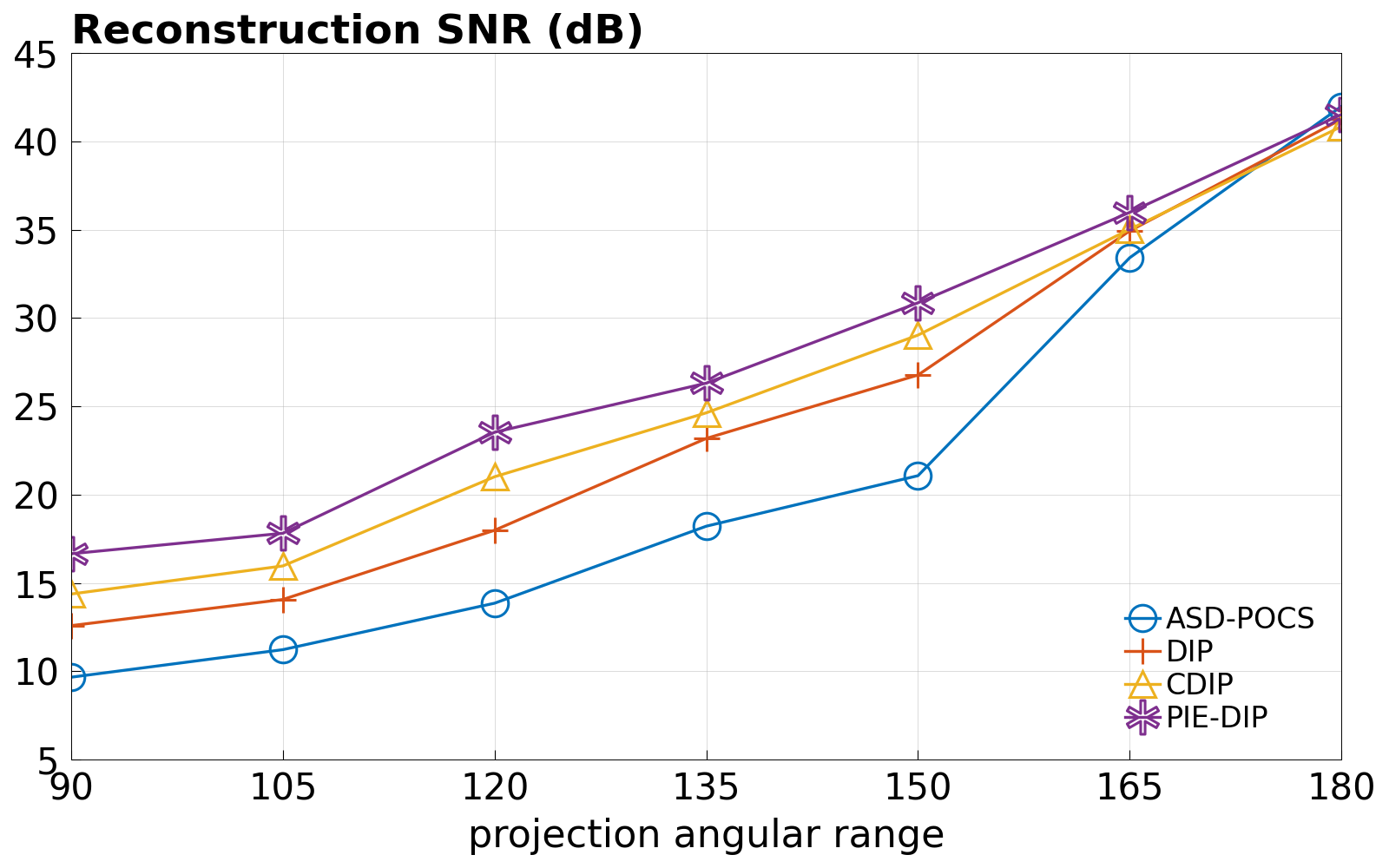}}
        \centerline{(b)}\medskip
    \end{minipage}
    \caption{Quantitative SNR comparison on the LIDC-IDRI dataset under (a) sparse-view and (b) limited-angle CT reconstruction settings. Each plotted value is the mean SNR over 100 patient cases randomly sampled directly from LIDC-IDRI. The sparse-view setting uses 30, 60, 90, 120, 180, and 360 projections over \(0^\circ\) to \(360^\circ\). The limited-angle setting uses angular ranges from \(90^\circ\) to \(180^\circ\) with two projections per degree.}
    \label{SVLARecon}
\end{figure}

Fig.~\ref{SVLARecon} compares ASD-POCS, DIP, CDIP, and PIE-DIP across acquisition levels. PIE-DIP achieves the highest mean SNR at all tested sparse-view levels and over limited-angle ranges of \(90^\circ\)--\(165^\circ\). Differences narrow as acquisition becomes more complete; ASD-POCS is comparable to CDIP at 360 projections and surpasses it at \(180^\circ\).

\subsection{Reconstruction Evaluation on Physical Object and SPARE Data}\label{sec:realCT}
\subsubsection{Finnish Inverse Problems Society Dataset}
We evaluate measured cheese and walnut CT data from the Finnish Inverse Problems Society~\cite{hamalainen2015tomographic,bubba2017tomographic}. Sparse-view reconstruction uses 30 projections uniformly distributed over \(0^\circ\)--\(360^\circ\); limited-angle reconstruction uses 60 projections over \(0^\circ\)--\(120^\circ\). The supplied high-resolution FBP reconstructions serve only as evaluation targets for SNR and SSIM.

Without matched reference scans, we manually construct coarse binary references that retain approximate boundaries and main internal structures but omit fine detail and local intensity variation. CDIP and PIE-DIP share the same reference for each object across both acquisition settings.

Pretrained models are omitted because matched training data are unavailable for these objects and acquisition systems. PICCS and NCPICCS are also omitted following poor preliminary results with these coarse binary references, which provide limited spatial and intensity compatibility for global reference-difference constraints.
\begin{figure*}
	\centering
	\begin{minipage}[t]{.19\linewidth}
		\vspace{0pt}
		\centering
		\centerline{\includegraphics[width=\linewidth]{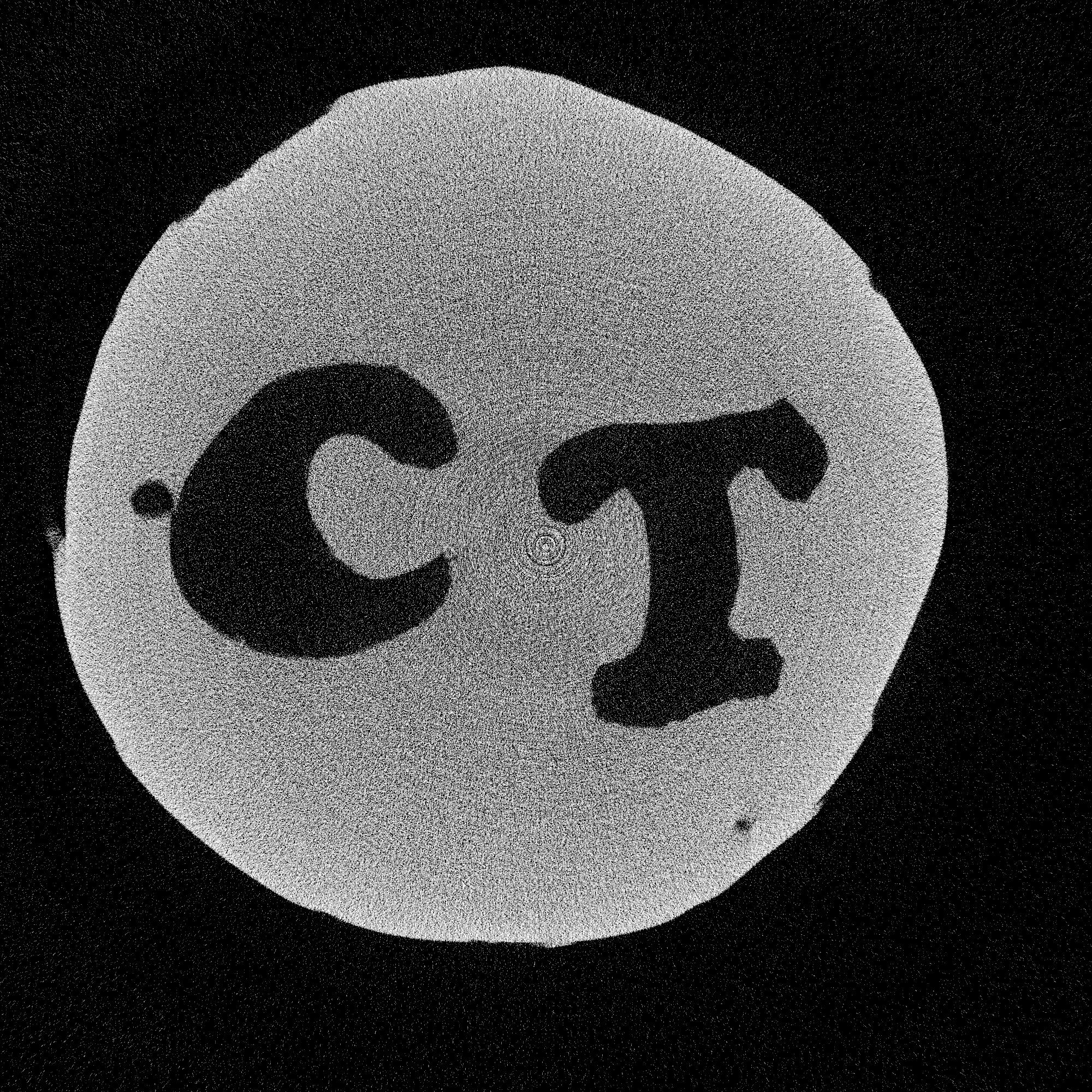}}\smallskip
		\centerline{\scriptsize\shortstack{\strut  High-resolution FBP\\\strut -/-}}\medskip
	\end{minipage}
	\begin{minipage}[t]{.19\linewidth}
		\vspace{0pt}
		\centering
		\centerline{\includegraphics[width=\linewidth]{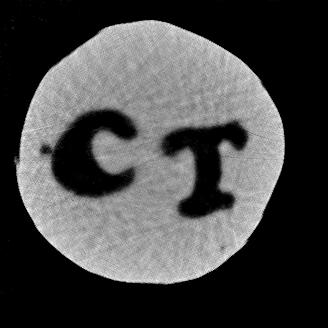}}\smallskip
		\centerline{\scriptsize\shortstack{\strut ASD-POCS\\\strut 19.21/0.686}}\medskip
	\end{minipage}
	\begin{minipage}[t]{.19\linewidth}
		\vspace{0pt}
		\centering
		\centerline{\includegraphics[width=\linewidth]{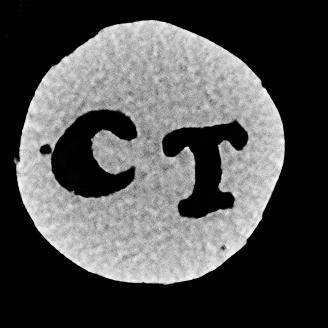}}\smallskip
		\centerline{\scriptsize\shortstack{\strut DIP\\\strut 19.61/0.682}}\medskip
	\end{minipage}
	\begin{minipage}[t]{.19\linewidth}
		\vspace{0pt}
		\centering
		\centerline{\includegraphics[width=\linewidth]{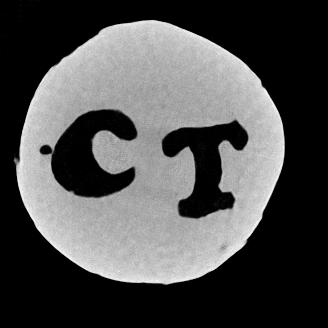}}\smallskip
		\centerline{\scriptsize\shortstack{\strut CDIP\\\strut 23.12/0.745}}\medskip
	\end{minipage}
	\begin{minipage}[t]{.19\linewidth}
		\vspace{0pt}
		\centering
		\centerline{\includegraphics[width=\linewidth]{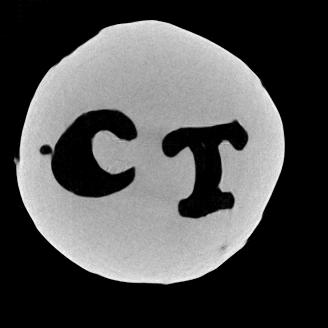}}\smallskip
		\centerline{\scriptsize\shortstack{\strut PIE-DIP\\\strut 24.85/0.781}}\medskip
	\end{minipage}

	\begin{minipage}[t]{.19\linewidth}
		\vspace{0pt}
		\centering
		\centerline{\includegraphics[width=\linewidth]{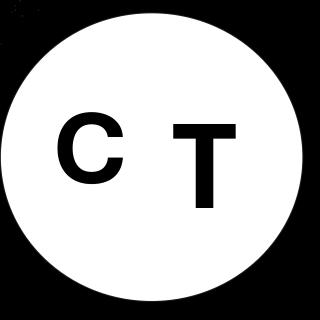}}\smallskip
		\centerline{\scriptsize \shortstack{\strut Reference\\\strut -/-}}\medskip
	\end{minipage}
	\begin{minipage}[t]{.19\linewidth}
		\vspace{0pt}
		\centering
		\centerline{\includegraphics[width=\linewidth]{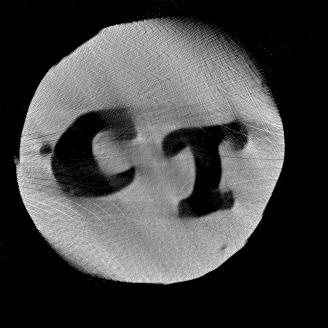}}\smallskip
		\centerline{\scriptsize\shortstack{\strut ASD-POCS\\\strut 14.79/0.621}}\medskip
	\end{minipage}
	\begin{minipage}[t]{.19\linewidth}
		\vspace{0pt}
		\centering
		\centerline{\includegraphics[width=\linewidth]{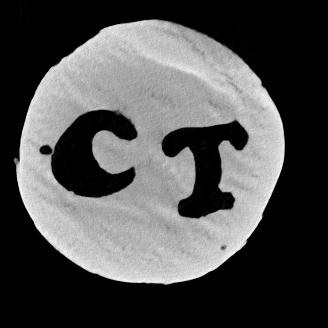}}\smallskip
		\centerline{\scriptsize\shortstack{\strut DIP\\\strut 19.91/0.731}}\medskip
	\end{minipage}
	\begin{minipage}[t]{.19\linewidth}
		\vspace{0pt}
		\centering
		\centerline{\includegraphics[width=\linewidth]{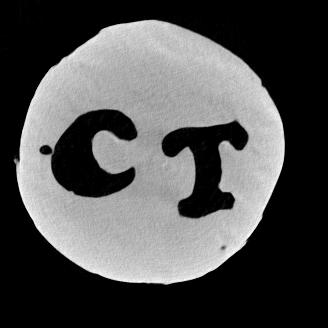}}\smallskip
		\centerline{\scriptsize\shortstack{\strut CDIP\\\strut 22.49/0.751}}\medskip
	\end{minipage}
	\begin{minipage}[t]{.19\linewidth}
		\vspace{0pt}
		\centering
		\centerline{\includegraphics[width=\linewidth]{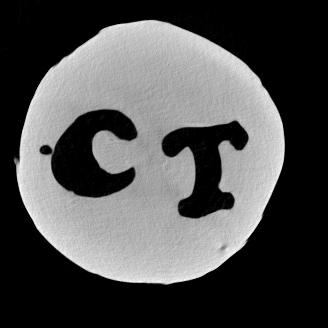}}\smallskip
		\centerline{\scriptsize\shortstack{\strut PIE-DIP\\\strut 23.25/0.768}}\medskip
	\end{minipage}
 	
	\begin{minipage}[t]{.19\linewidth}
		\vspace{0pt}
		\centering
		\centerline{\includegraphics[width=\linewidth]{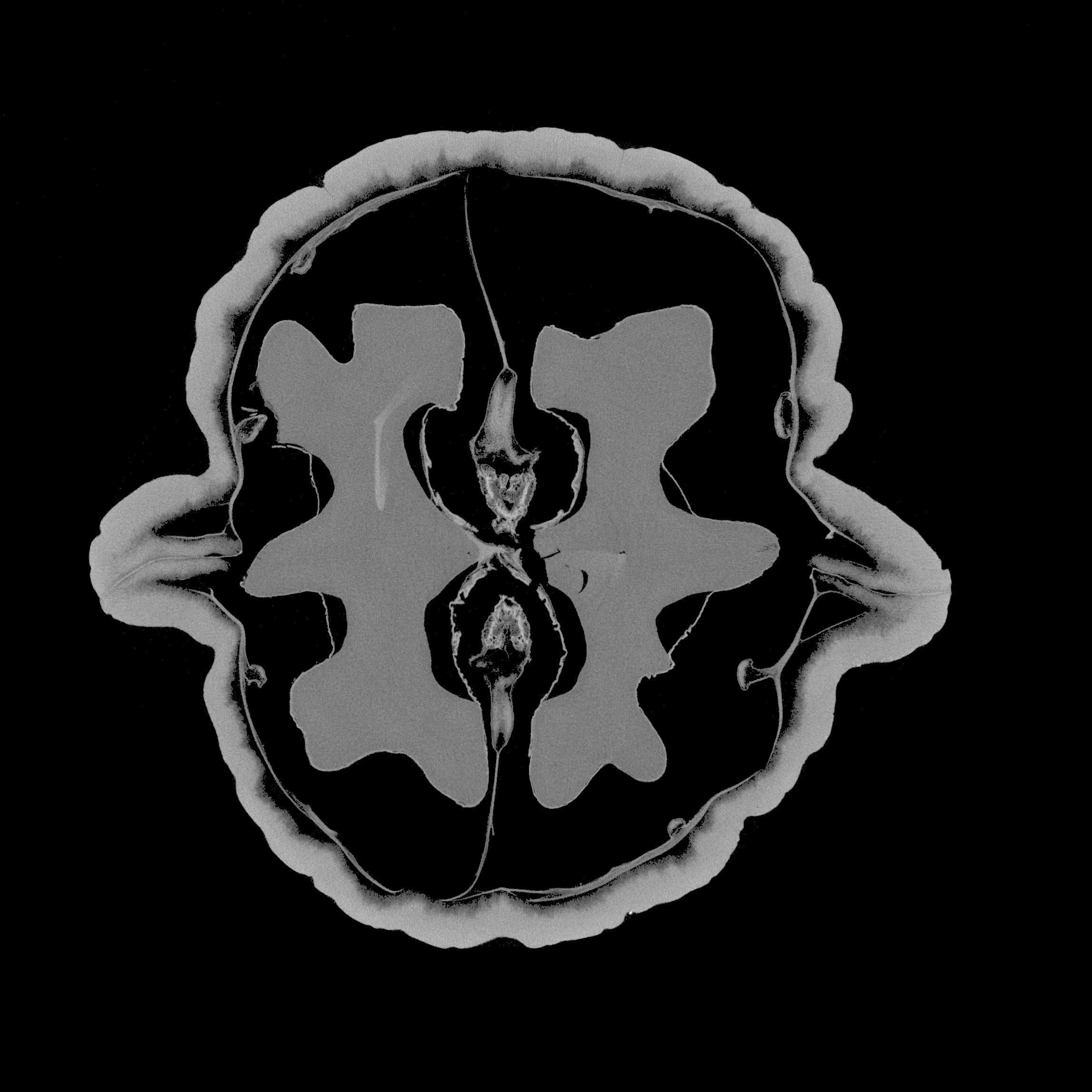}}\smallskip
		\centerline{\scriptsize \shortstack{\strut High-resolution FBP\\\strut -/-}}\medskip
	\end{minipage}
	\begin{minipage}[t]{.19\linewidth}
		\vspace{0pt}
		\centering
		\centerline{\includegraphics[width=\linewidth]{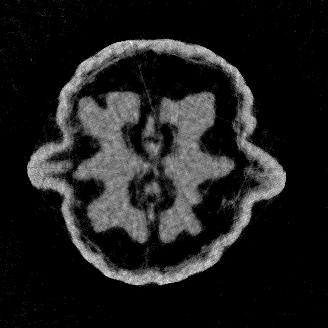}}\smallskip
		\centerline{\scriptsize\shortstack{\strut ASD-POCS\\\strut 11.61/0.738}}\medskip
	\end{minipage}
	\begin{minipage}[t]{.19\linewidth}
		\vspace{0pt}
		\centering
		\centerline{\includegraphics[width=\linewidth]{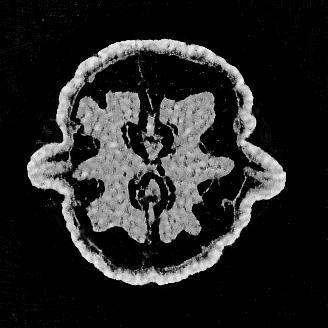}}\smallskip
		\centerline{\scriptsize\shortstack{\strut DIP\\\strut 12.35/0.745}}\medskip
	\end{minipage}
	\begin{minipage}[t]{.19\linewidth}
		\vspace{0pt}
		\centering
		\centerline{\includegraphics[width=\linewidth]{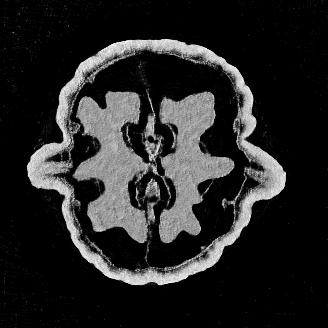}}\smallskip
		\centerline{\scriptsize\shortstack{\strut CDIP\\\strut 13.46/0.751}}\medskip
	\end{minipage}
	\begin{minipage}[t]{.19\linewidth}
		\vspace{0pt}
		\centering
		\centerline{\includegraphics[width=\linewidth]{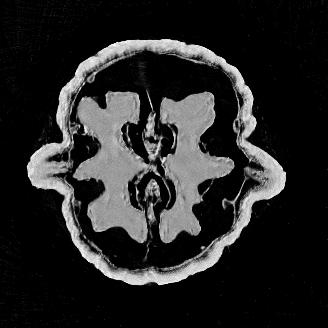}}\smallskip
		\centerline{\scriptsize\shortstack{\strut PIE-DIP\\\strut 14.25/0.785}}\medskip
	\end{minipage}
    
	\begin{minipage}[t]{.19\linewidth}
		\vspace{0pt}
		\centering
		\centerline{\includegraphics[width=\linewidth]{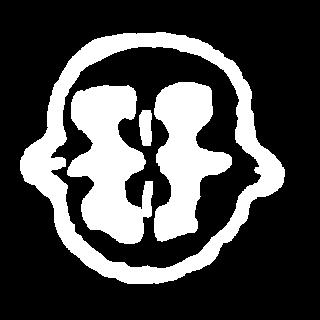}}\smallskip
		\centerline{\scriptsize\shortstack{\strut  Reference\\\strut -/-}}\medskip
	\end{minipage}
	\begin{minipage}[t]{.19\linewidth}
		\vspace{0pt}
		\centering
		\centerline{\includegraphics[width=\linewidth]{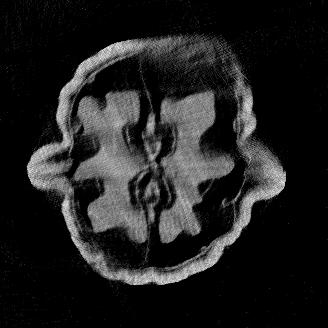}}\smallskip
		\centerline{\scriptsize\shortstack{\strut ASD-POCS\\\strut 9.41/0.671}}\medskip
	\end{minipage}
	\begin{minipage}[t]{.19\linewidth}
		\vspace{0pt}
		\centering
		\centerline{\includegraphics[width=\linewidth]{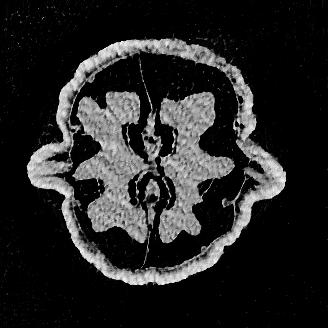}}\smallskip
		\centerline{\scriptsize\shortstack{\strut DIP\\\strut 11.77/0.680}}\medskip
	\end{minipage}
	\begin{minipage}[t]{.19\linewidth}
		\vspace{0pt}
		\centering
		\centerline{\includegraphics[width=\linewidth]{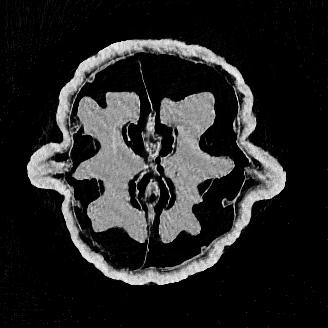}}\smallskip
		\centerline{\scriptsize\shortstack{\strut CDIP\\\strut 13.49/0.738}}\medskip
	\end{minipage}
	\begin{minipage}[t]{.19\linewidth}
		\vspace{0pt}
		\centering
		\centerline{\includegraphics[width=\linewidth]{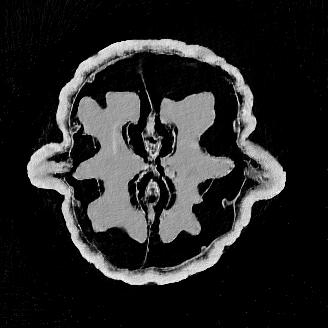}}\smallskip
		\centerline{\scriptsize\shortstack{\strut PIE-DIP\\\strut 14.59/0.752}}\medskip
	\end{minipage}
	\caption{Fan-beam reconstruction results for a carved cheese and a walnut from the Finnish Inverse Problems Society dataset. The first two rows show the cheese under sparse-view and limited-angle settings, respectively, and the last two rows show the corresponding walnut results. Sparse-view reconstruction uses 30 projections uniformly distributed over $0^{\circ}$--$360^{\circ}$, whereas limited-angle reconstruction uses 60 projections over $0^{\circ}$--$120^{\circ}$. The high-resolution FBP images shown in the first and third rows serve only as the evaluation targets. The manually constructed images shown in the second and fourth rows are the structural references used by CDIP and PIE-DIP for the corresponding objects. The remaining columns show ASD-POCS, DIP, CDIP, and PIE-DIP. The reported values are SNR (dB)/SSIM.}
	\label{reconresult3}
\end{figure*}

PIE-DIP achieves the highest SNR and SSIM among the compared methods in all four tests (Fig.~\ref{reconresult3}). Relative to ASD-POCS, standard DIP, and CDIP, respectively, SNR gains for cheese are 5.64, 5.24, and 1.73~dB under sparse views and 8.46, 3.34, and 0.76~dB under limited angles. For walnut, the corresponding gains are 2.64, 1.90, and 0.79~dB under sparse views and 5.18, 2.82, and 1.10~dB under limited angles.

These results support combining measurement-derived input with local patch guidance to exploit coarse structural references.

\subsubsection{SPARE Challenge Dataset}\label{sec:spare_singlephase}
The SPARE Challenge benchmarks sparse-view 4D-CBCT reconstruction from one-minute scans, including patient-based MC simulations and clinical Varian half-fan acquisitions~\cite{shieh2019spare}. MC ground truth comes from patient 4D-CT images, whereas clinical reference standards are reconstructed from oversampled acquisitions. Respiratory binning leaves few, unevenly distributed projections per phase.

We use all 32 MC scans from 12 patients and all 30 clinical Varian scans from 5 patients. For each scan, we perform single-phase reconstruction using the original MC simulated or clinical Varian raw projections with the supplied geometry and phase assignments.

Phase-specific Feldkamp--Davis--Kress (FDK) reconstruction serves as the baseline. Phase FDK and all DIP-based methods use the same selected-phase projections. Full FDK uses all available projections of each scan before respiratory binning and provides the motion-affected prior for CDIP and PIE-DIP. DIP-based methods use the common U-Net, 5000-iteration budget, and optimization settings; PIE-DIP also uses the common patch-prior parameters (Section~\ref{sec:Proposed Methods}).

Metrics are evaluated within the supplied body mask using normalized floating-point images. Table~\ref{tab:spare_fdk_improvement} reports the mean and sample standard deviation of paired per-scan metric differences relative to phase FDK across all scans in each dataset. SNR gains are in dB, and SSIM gains are absolute differences.
\begin{table}[htbp]
    \centering
    \caption{Improvement relative to phase FDK in the SPARE single-phase experiments using the complete MC simulated projection and clinical Varian raw projection datasets. Values are mean $\pm$ sample standard deviation of the per-scan metric differences over all 32 MC scans and 30 clinical Varian scans. SNR improvements are in dB; SSIM improvements are absolute differences. Metrics are evaluated within the supplied body mask using the original floating-point images. Full FDK is used as the prior image by CDIP and PIE-DIP; it is not the baseline used to calculate these improvements.}
    \label{tab:spare_fdk_improvement}
    \begingroup
    \small
    \renewcommand{\arraystretch}{1.15}
    \begin{tabular}{llcc}
        \toprule
        Subset & Method & $\Delta$SNR (dB) & $\Delta$SSIM \\
        \midrule
        MC (32 scans) & CDIP & $15.16 \pm 2.75$ & $0.588 \pm 0.056$ \\
         & DIP & $8.18 \pm 2.78$ & $0.357 \pm 0.058$ \\
         & PIE-DIP & $\mathbf{17.13 \pm 2.31}$ & $\mathbf{0.638 \pm 0.035}$ \\
        \midrule
        Varian (30 scans) & CDIP & $18.11 \pm 3.83$ & $0.643 \pm 0.210$ \\
         & DIP & $14.71 \pm 3.37$ & $0.514 \pm 0.236$ \\
         & PIE-DIP & $\mathbf{19.35 \pm 2.98}$ & $\mathbf{0.682 \pm 0.165}$ \\
        \bottomrule
    \end{tabular}
    \endgroup
\end{table}

\begin{figure*}
    \centering
    \begin{minipage}[b]{0.158\linewidth}
        \centering
        \centerline{\includegraphics[width=\linewidth]{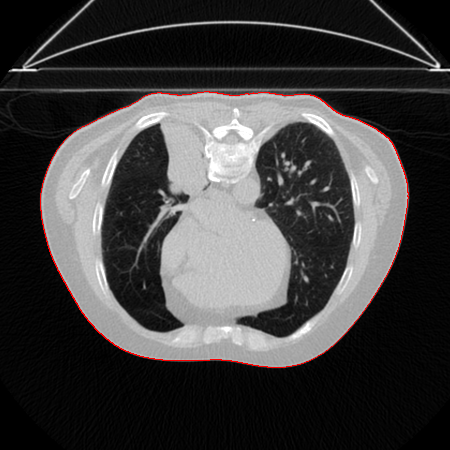}}\smallskip
        \centerline{\scriptsize\shortstack{Ground truth\\-/-}}\medskip
    \end{minipage}\hfill
    \begin{minipage}[b]{0.158\linewidth}
        \centering
        \centerline{\includegraphics[width=\linewidth]{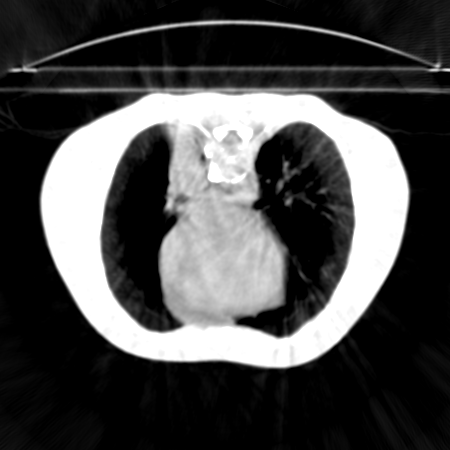}}\smallskip
        \centerline{\scriptsize\shortstack{Full FDK\\6.89/0.560}}\medskip
    \end{minipage}\hfill
    \begin{minipage}[b]{0.158\linewidth}
        \centering
        \centerline{\includegraphics[width=\linewidth]{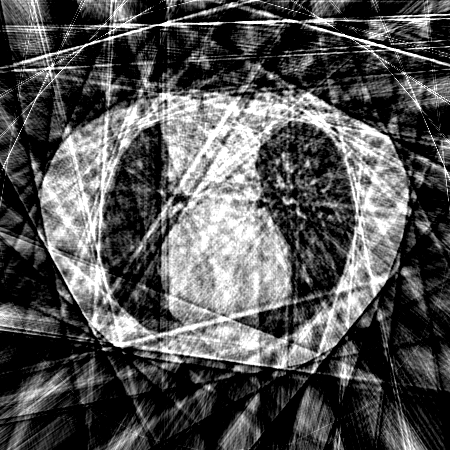}}\smallskip
        \centerline{\scriptsize\shortstack{Phase FDK\\8.90/0.228}}\medskip
    \end{minipage}\hfill
    \begin{minipage}[b]{0.158\linewidth}
        \centering
        \centerline{\includegraphics[width=\linewidth]{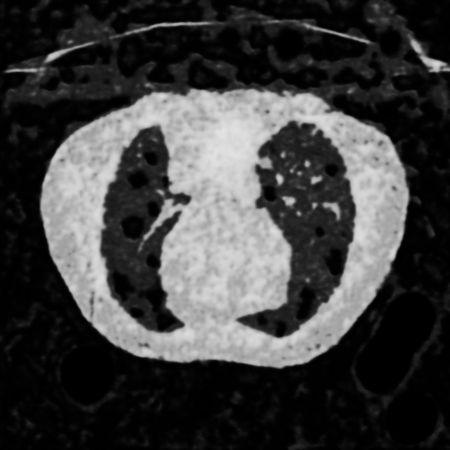}}\smallskip
        \centerline{\scriptsize\shortstack{DIP\\16.92/0.557}}\medskip
    \end{minipage}\hfill
    \begin{minipage}[b]{0.158\linewidth}
        \centering
        \centerline{\includegraphics[width=\linewidth]{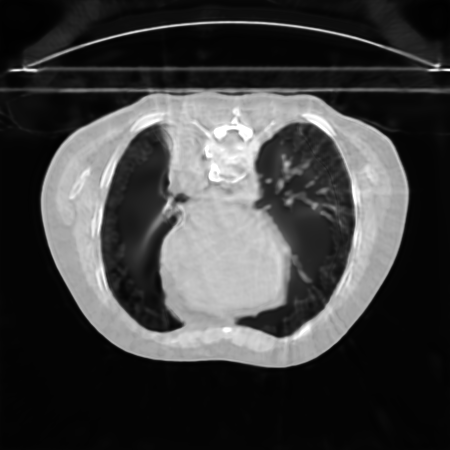}}\smallskip
        \centerline{\scriptsize\shortstack{CDIP\\20.97/0.749}}\medskip
    \end{minipage}\hfill
    \begin{minipage}[b]{0.158\linewidth}
        \centering
        \centerline{\includegraphics[width=\linewidth]{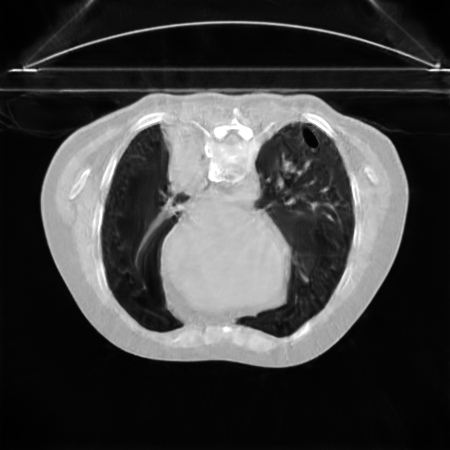}}\smallskip
        \centerline{\scriptsize\shortstack{PIE-DIP\\23.47/0.823}}\medskip
    \end{minipage}\par
    \medskip
    \begin{minipage}[b]{0.158\linewidth}
        \centering
        \centerline{\includegraphics[width=\linewidth]{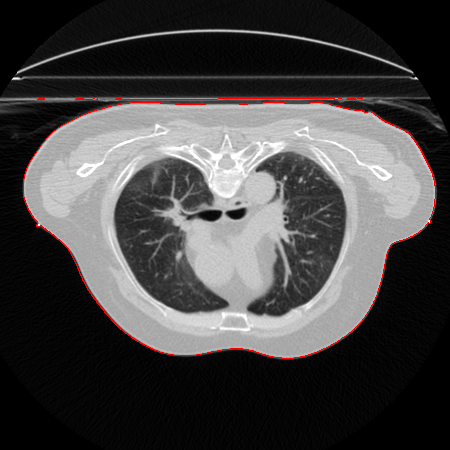}}\smallskip
        \centerline{\scriptsize\shortstack{Ground truth\\-/-}}\medskip
    \end{minipage}\hfill
    \begin{minipage}[b]{0.158\linewidth}
        \centering
        \centerline{\includegraphics[width=\linewidth]{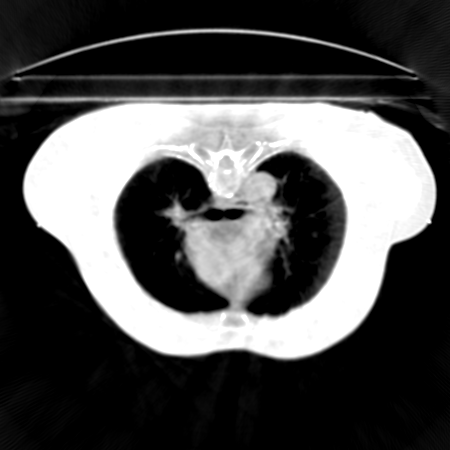}}\smallskip
        \centerline{\scriptsize\shortstack{Full FDK\\6.78/0.658}}\medskip
    \end{minipage}\hfill
    \begin{minipage}[b]{0.158\linewidth}
        \centering
        \centerline{\includegraphics[width=\linewidth]{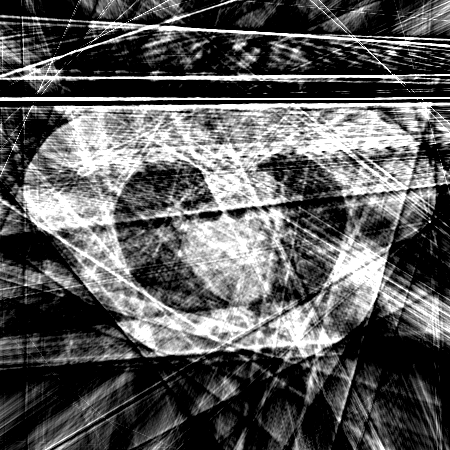}}\smallskip
        \centerline{\scriptsize\shortstack{Phase FDK\\3.67/0.253}}\medskip
    \end{minipage}\hfill
    \begin{minipage}[b]{0.158\linewidth}
        \centering
        \centerline{\includegraphics[width=\linewidth]{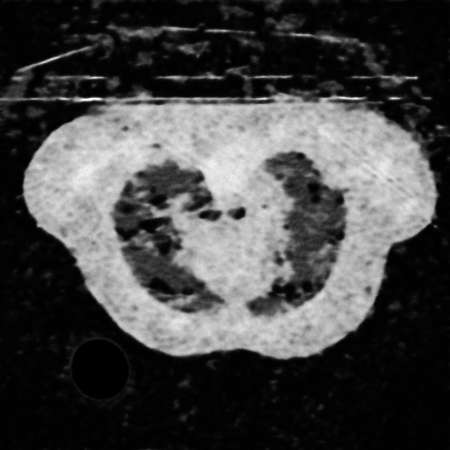}}\smallskip
        \centerline{\scriptsize\shortstack{DIP\\15.86/0.665}}\medskip
    \end{minipage}\hfill
    \begin{minipage}[b]{0.158\linewidth}
        \centering
        \centerline{\includegraphics[width=\linewidth]{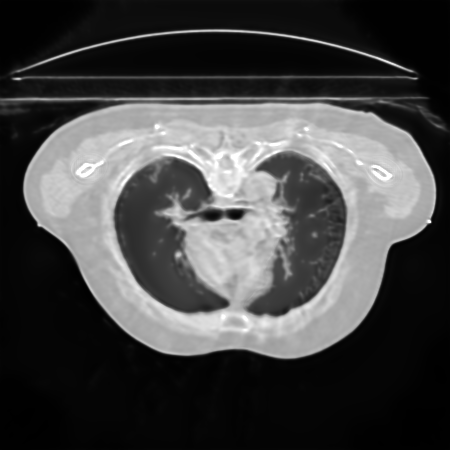}}\smallskip
        \centerline{\scriptsize\shortstack{CDIP\\22.27/0.871}}\medskip
    \end{minipage}\hfill
    \begin{minipage}[b]{0.158\linewidth}
        \centering
        \centerline{\includegraphics[width=\linewidth]{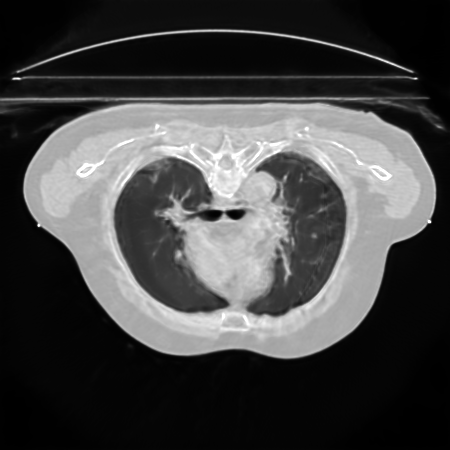}}\smallskip
        \centerline{\scriptsize\shortstack{PIE-DIP\\24.12/0.902}}\medskip
    \end{minipage}\par
    \medskip
    \begin{minipage}[b]{0.158\linewidth}
        \centering
        \centerline{\includegraphics[width=\linewidth]{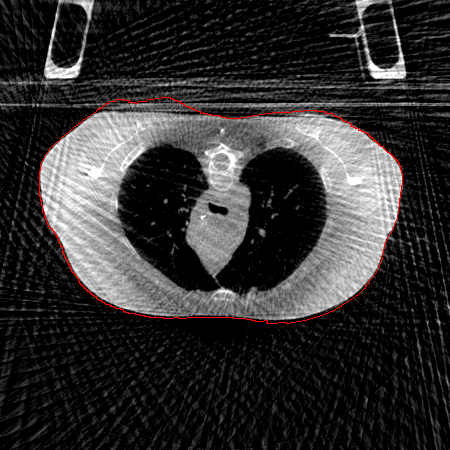}}\smallskip
        \centerline{\scriptsize\shortstack{Ground truth\\-/-}}\medskip
    \end{minipage}\hfill
    \begin{minipage}[b]{0.158\linewidth}
        \centering
        \centerline{\includegraphics[width=\linewidth]{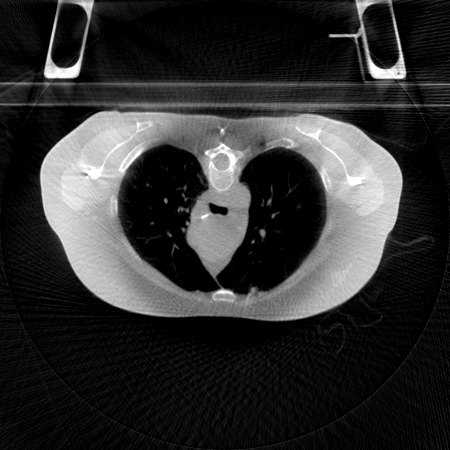}}\smallskip
        \centerline{\scriptsize\shortstack{Full FDK\\17.29/0.662}}\medskip
    \end{minipage}\hfill
    \begin{minipage}[b]{0.158\linewidth}
        \centering
        \centerline{\includegraphics[width=\linewidth]{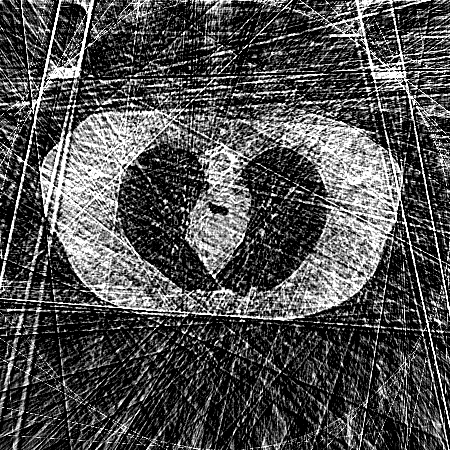}}\smallskip
        \centerline{\scriptsize\shortstack{Phase FDK\\3.50/0.206}}\medskip
    \end{minipage}\hfill
    \begin{minipage}[b]{0.158\linewidth}
        \centering
        \centerline{\includegraphics[width=\linewidth]{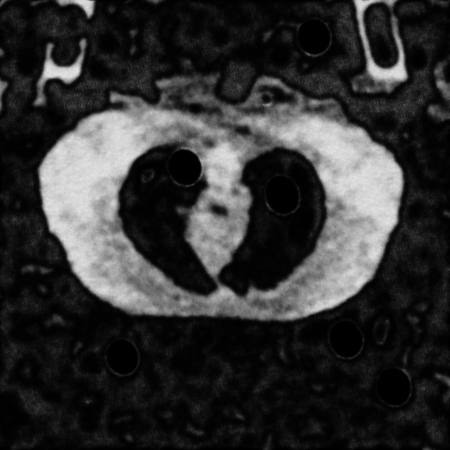}}\smallskip
        \centerline{\scriptsize\shortstack{DIP\\14.20/0.479}}\medskip
    \end{minipage}\hfill
    \begin{minipage}[b]{0.158\linewidth}
        \centering
        \centerline{\includegraphics[width=\linewidth]{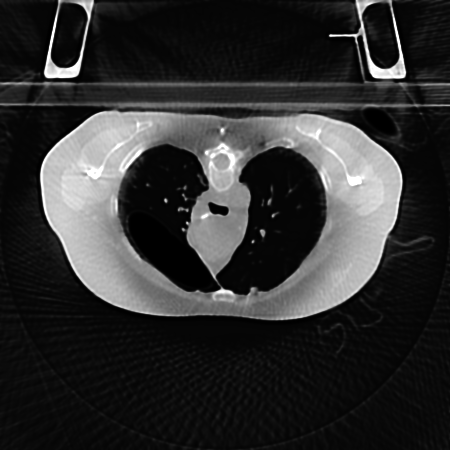}}\smallskip
        \centerline{\scriptsize\shortstack{CDIP\\17.13/0.609}}\medskip
    \end{minipage}\hfill
    \begin{minipage}[b]{0.158\linewidth}
        \centering
        \centerline{\includegraphics[width=\linewidth]{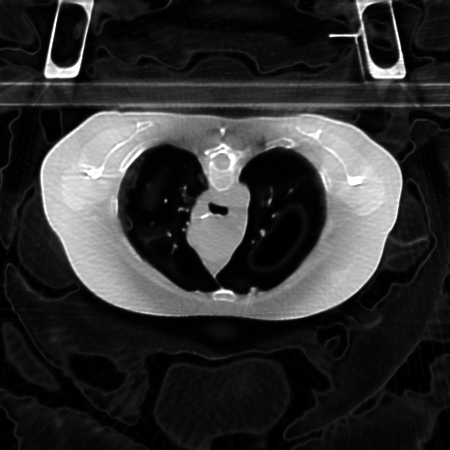}}\smallskip
        \centerline{\scriptsize\shortstack{PIE-DIP\\18.02/0.689}}\medskip
    \end{minipage}\par
    \medskip
    \begin{minipage}[b]{0.158\linewidth}
        \centering
        \centerline{\includegraphics[width=\linewidth]{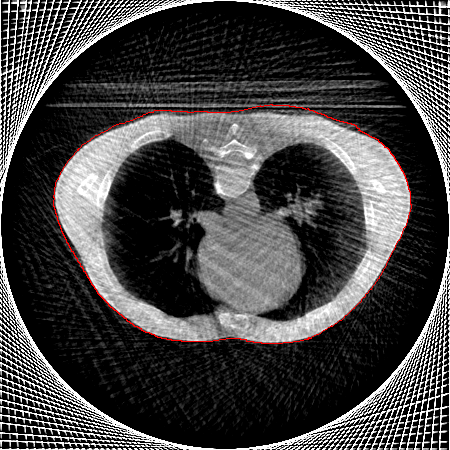}}\smallskip
        \centerline{\scriptsize\shortstack{Ground truth\\-/-}}\medskip
    \end{minipage}\hfill
    \begin{minipage}[b]{0.158\linewidth}
        \centering
        \centerline{\includegraphics[width=\linewidth]{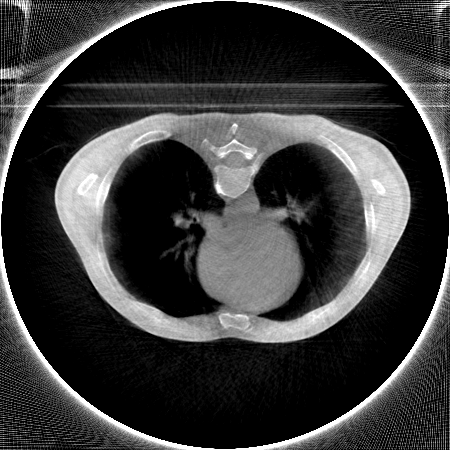}}\smallskip
        \centerline{\scriptsize\shortstack{Full FDK\\16.77/0.882}}\medskip
    \end{minipage}\hfill
    \begin{minipage}[b]{0.158\linewidth}
        \centering
        \centerline{\includegraphics[width=\linewidth]{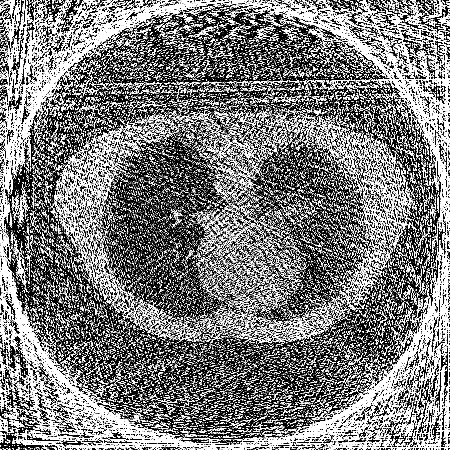}}\smallskip
        \centerline{\scriptsize\shortstack{Phase FDK\\-3.67/0.091}}\medskip
    \end{minipage}\hfill
    \begin{minipage}[b]{0.158\linewidth}
        \centering
        \centerline{\includegraphics[width=\linewidth]{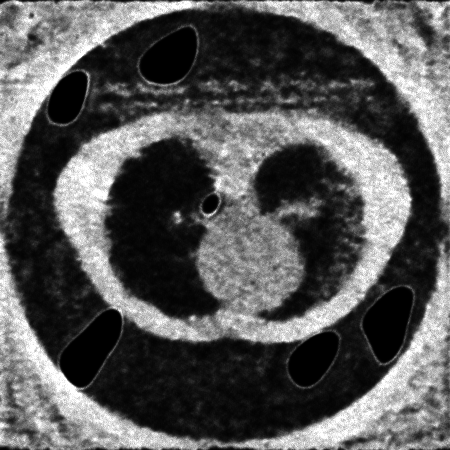}}\smallskip
        \centerline{\scriptsize\shortstack{DIP\\13.18/0.785}}\medskip
    \end{minipage}\hfill
    \begin{minipage}[b]{0.158\linewidth}
        \centering
        \centerline{\includegraphics[width=\linewidth]{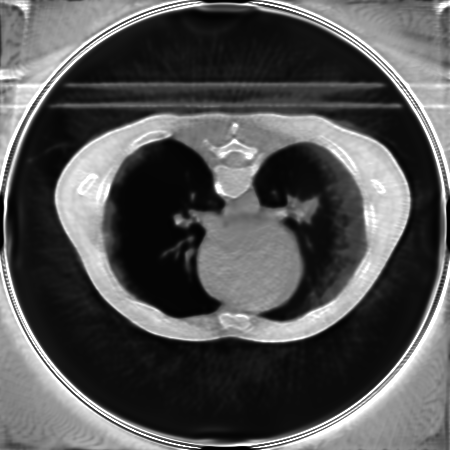}}\smallskip
        \centerline{\scriptsize\shortstack{CDIP\\16.94/0.888}}\medskip
    \end{minipage}\hfill
    \begin{minipage}[b]{0.158\linewidth}
        \centering
        \centerline{\includegraphics[width=\linewidth]{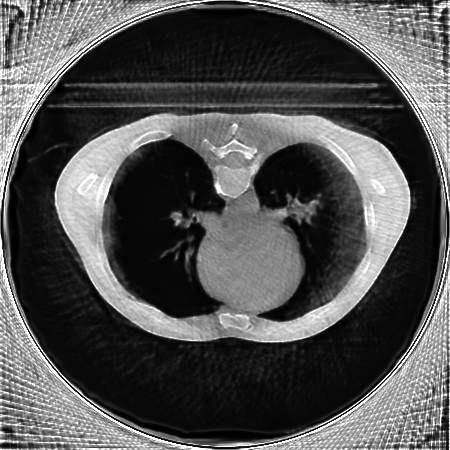}}\smallskip
        \centerline{\scriptsize\shortstack{PIE-DIP\\17.94/0.906}}\medskip
    \end{minipage}\par
    \caption{Representative single-phase few-view reconstructions for two cases from the MC simulation dataset (first two rows) and two cases from the clinical Varian dataset (last two rows). From left to right, the columns show ground truth with the body-mask outline, full FDK, phase FDK, DIP, CDIP, and PIE-DIP. The red outlines in the first column indicate the body regions used for quantitative evaluation. For the clinical Varian cases, ground truth denotes the supplied reference reconstruction. Full FDK provides the prior image for CDIP and PIE-DIP; phase FDK and the iterative methods use the same phase-specific projections. Values below the images are SNR (dB)/SSIM within the body mask. All images in each row share the same display window.}
    \label{fig:spare_singlephase}
\end{figure*}

All three DIP-based methods improve mean SNR and SSIM over phase FDK. PIE-DIP achieves the largest mean improvements in both datasets and the highest SNR and SSIM among the displayed methods in the four representative cases (Fig.~\ref{fig:spare_singlephase}).

\section{Discussion}\label{sec:diss}
The ablation studies support complementary roles for network input design and local patch guidance. An image already fitted to the measurements can have a small projection residual yet retain artifacts, which a flexible network may reproduce. Modified back-projection instead supplies object-specific linear structure without iterative data fitting, complementing the nonlinear implicit prior of DIP. Random modulation is intended to weaken reliance on coherent streak patterns. The gains in Fig.~\ref{netinput} support this input construction, although severe artifacts in a modulating reference can still limit its benefit.

The patch prior provides a different way to use structural information. Position-independent matching allows local reference structures to contribute without requiring pixel-wise correspondence. Within a matched group, collaborative filtering can retain shared structure and attenuate weak transform coefficients associated with inconsistent artifact components, providing a rationale for suppressing directional streaks. The gains with the less similar Reference 2 support this local use of reference information. Spatial independence applies to patch matching; an optional reference can still modulate the fixed network input.

Fig.~\ref{patchprior_ablation} supports combining reference and reconstruction patches rather than relying on either source alone. The scheduled mixture achieves the highest SNR and SSIM in both reference conditions, exceeding BM3D and both single-source dictionaries. A plausible explanation is that reference patches support unreliable early reconstructions, whereas later reconstruction patches allow case-specific refinement. The improvement from Fig.~\ref{netinput}(i) to Fig.~\ref{netinput}(m) separately supports patch guidance without any external reference.

The optimization curves distinguish projection fitting from reconstruction accuracy. The data-fidelity loss in Fig.~\ref{trainingperformance} excludes the patch term, and its ordering across methods is opposite to that of SNR. Under incomplete acquisition, weakly constrained image components can contribute substantial error without a large projection residual. The results are therefore consistent with a regularizing effect of the proposed guidance. The overall decline in loss and late SNR plateau also provide empirical evidence of stable reconstruction accuracy under the tested optimization settings.

The acquisition-level results show where additional guidance is most valuable: differences narrow as more views or wider angular coverage become available (Fig.~\ref{SVLARecon}). This is consistent with stronger measurements reducing the ambiguity left to image priors. In the representative task-matched examples (Fig.~\ref{reconresult}), PIE-DIP also achieves competitive reconstruction quality without offline training, whereas CvG-Diff and DOLCE each use images from 500 patients for training. This makes PIE-DIP an option when matched training data are limited.

The additional datasets test distinct forms of imperfect reference information. Finnish Inverse Problems Society data show that coarse binary references remain useful despite missing fine details. SPARE extends evaluation to all MC simulated and clinical Varian raw projection scans, with the same motion-affected full FDK prior supplied to CDIP and PIE-DIP. In these single-phase tests, PIE-DIP achieves the largest mean SNR and SSIM gains over phase FDK among the compared DIP-based methods in both subsets (Table~\ref{tab:spare_fdk_improvement}). Together, these results support the combined design with both coarse and acquisition-derived references.

Periodic updates avoid repeating patch matching at every iteration. During each update, the network continues optimization with current guidance while the next dictionary and guiding patches are generated in parallel. If generation finishes within the update interval, the network need not wait. The reported GPU measurements show almost no increase in average iteration time.

Patch reliability remains an important limitation. Normalized grayscale distances can match artifacts as well as true structures, while a shared guidance weight does not distinguish match reliability. Future work will explore matching criteria that explicitly combine intensity and local structure, and use within-group consistency to adjust guidance strength across image regions. The CT platform under development will support further evaluation with sparse-view, limited-angle, and inverse-geometry acquisitions.

\section{Conclusion}\label{sec:conclusion}
PIE-DIP combines a measurement-derived network input with collaborative patch-group guidance for sparse-view and limited-angle CT reconstruction without offline training. The modified back-projection input introduces a linear structural prior that complements DIP's nonlinear implicit prior. The patch constraint uses reconstruction self-similarity and optional reference structures, while a scheduled mixed dictionary progressively shifts toward reconstruction patches.

Ablation studies support the complementary contributions of the two components. Experiments on the FORBILD phantom, LIDC-IDRI patient-image simulations, Finnish Inverse Problems Society measured data, and single-phase SPARE Challenge data demonstrate improved reconstruction under the tested incomplete-acquisition and imperfect-reference conditions. These findings highlight the value of combining measurement-derived structure with selective local guidance when projection information is limited.

\bibliographystyle{medphy}
\bibliography{refs}

\end{document}